\documentclass[]{aa} 
\usepackage[]{natbib}
\usepackage{graphicx,color}
\usepackage{txfonts}
\usepackage[colorlinks=true, citecolor=blue]{hyperref}

\newcommand{\orcid}[1]{\href{https://orcid.org/#1}{\includegraphics[width=10pt]{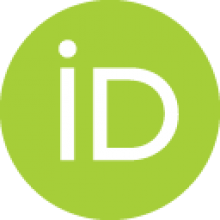}}}

\begin{document} 

\title{APOGEE spectroscopic evidence for chemical anomalies in dwarf galaxies: The case of M~54 and Sagittarius}
                      
	\author{
		Jos\'e G. Fern\'andez-Trincado\inst{1,2,3}\thanks{To whom correspondence should be addressed; E-mail: jose.fernandez@uda.cl and/or jfernandezt87@gmail.com}\orcid{0000-0003-3526-5052}, 
		Timothy C. Beers\inst{4}\orcid{0000-0003-4573-6233},
		Dante Minniti\inst{5,6}\orcid{0000-0002-7064-099X},
		Christian Moni Bidin\inst{7},
		Beatriz Barbuy\inst{8}, 
		Sandro Villanova\inst{9},
		Doug Geisler\inst{9,10,11},
		Richard R. Lane\inst{2}, 
		Alexandre Roman-Lopes\inst{10}\orcid{0000-0002-1379-4204}
		\and
		Dmitry Bizyaev\inst{12, 13}
		}
	
	\authorrunning{Jos\'e G. Fern\'andez-Trincado et al.} 
	
\institute{
	    Institut Utinam, CNRS UMR 6213, Universit\'e Bourgogne-Franche-Comt\'e, OSU THETA Franche-Comt\'e, Observatoire de Besan\c{c}on, \\ BP 1615, 25010 Besan\c{c}on Cedex, France
	    \and
		Instituto de Astronom\'ia y Ciencias Planetarias, Universidad de Atacama, Copayapu 485, Copiap\'o, Chile
		\and 
		Centro de Investigaci\'on en Astronom\'ia, Universidad Bernardo O Higgins, Avenida Viel 1497, Santiago, Chile
		\and
		Department of Physics and JINA Center for the Evolution of the Elements, University of Notre Dame, Notre Dame, IN 46556, USA
		\and 
		Depto. de Cs. F\'isicas, Facultad de Ciencias Exactas, Universidad Andres Bello, Av. Fern\'andez Concha 700, Las Condes, Santiago, Chile
		\and
		Vatican Observatory, V00120 Vatican City State, Italy
		\and 
		Instituto de Astronom\'ia, Universidad Cat\'olica del Norte, Av. Angamos 0610, Antofagasta, Chile
		\and 
		Universidade de S\~ao Paulo, IAG, Rua do Mat\~ao 1226, Cidade Universit\'aria, S\~ao Paulo 05508-900, Brazil
		\and
		Departamento de Astronom\'ia, Casilla 160-C, Universidad de Concepci\'on, Concepci\'on, Chile
		\and
		Departamento de Astronom\'ia, Universidad de La Serena, 1700000 La Serena, Chile
		\and
		Instituto de Investigaci\'on Multidisciplinario en Ciencia y Tecnolog\'ia, Universidad de La Serena. Benavente 980, La Serena, Chile
		\and
       Apache Point Observatory and New Mexico State University, Sunspot, NM, 88349, USA
      \and
        Sternberg Astronomical Institute, Moscow State University, Moscow, Russia 
    }
	
	\date{Received ...; Accepted ...}
	\titlerunning{Globular cluster disruption in dwarf galaxies}
	
	
	\abstract
	{
		We present evidence for globular cluster stellar debris in a dwarf galaxy system (Sagittarius: Sgr) based on an analysis of high-resolution \textit{H}-band spectra from the Apache Point Observatory Galactic Evolution Experiment (APOGEE) survey. We add [N/Fe], [Ti/Fe], and [Ni/Fe] abundance ratios to the existing sample of potential members of M~54; this is the first time that [N/Fe] abundances are derived for a large number of stars in M~54. Our study reveals the existence of a significant population of nitrogen- (with a large spread, $\gtrsim1$ dex) and aluminum-enriched stars with moderate Mg depletion in the core of the M~54$+$Sagittarius system, which shares the light element anomalies characteristic of second-generation globular cluster stars (GCs), thus tracing the typical phenomenon of multiple stellar populations seen in other Galactic GCs at similar metallicity, confirming earlier results based on the Na-O anti-correlation. We further show that most of the stars in M~54 exhibit different chemical - patterns evidently not present in Sgr field stars. Furthermore, we report the serendipitous discovery of a nitrogen-enhanced extra-tidal star with GC second-generation-like chemical patterns for which both chemical and kinematic evidence is commensurate with the hypothesis that the star has been ejected from M~54. Our findings support the existence of chemical anomalies associated with likely tidally shredded GCs in dwarf galaxies in the Local Group and motivate future searches for such bonafide stars along other known Milky Way streams. 
	}
	
	\maketitle

	\section{Introduction}
	\label{section1}
	
\begin{figure}	
	\begin{center}
		\includegraphics[width=90mm]{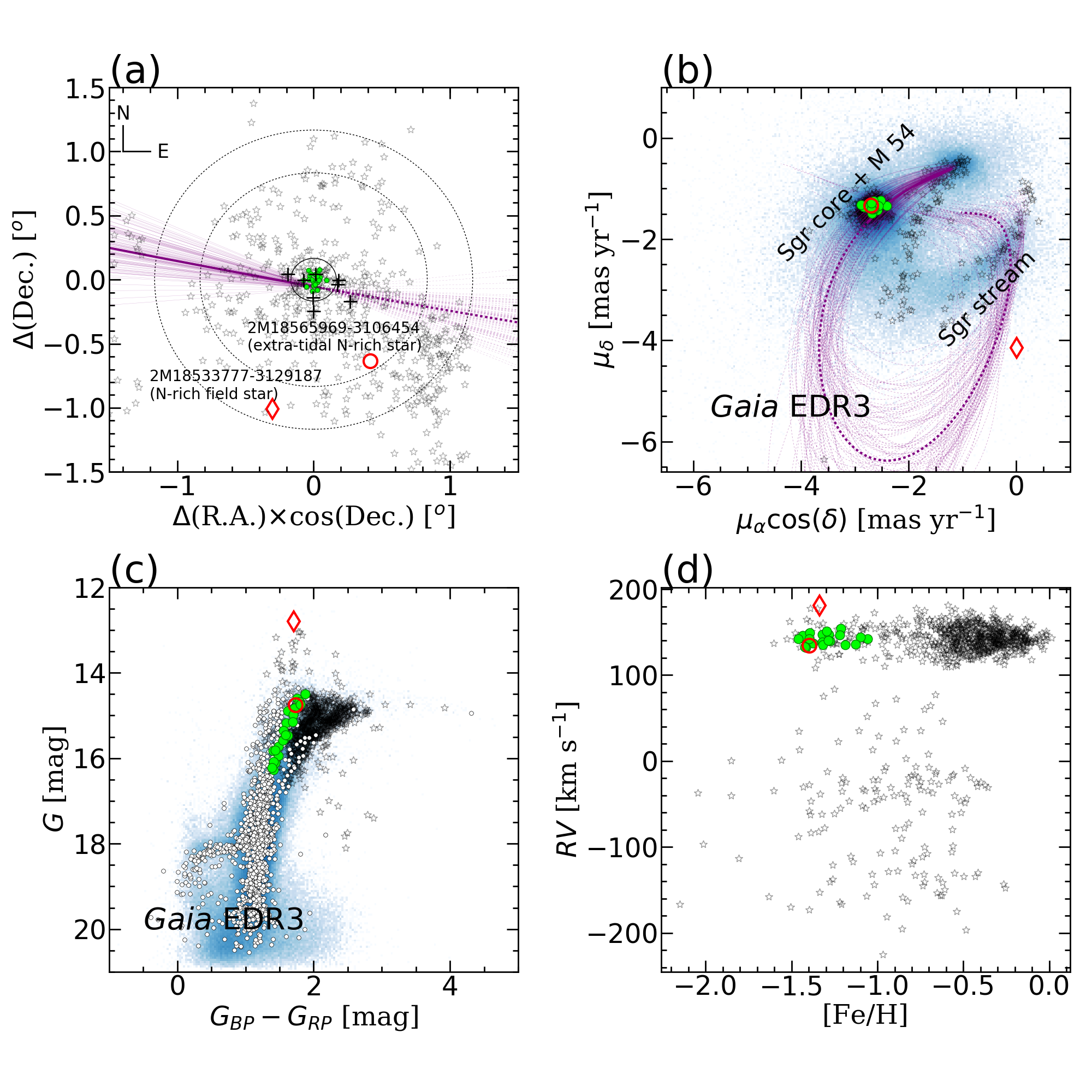}
		\caption{Panel (a): Spatial positions of the stars in our sample, with the tidal radius ($r_t=10$\arcmin) of M~54 over-plotted with a solid line. The open red symbols designate N-rich stars (the diamond symbol refers to a field star, while the open circle highlights the extra-tidal member of M~54). The lime circles designate the M~54 population analyzed in this work, while the black plus symbols designate the stars analyzed by \citet{Nataf2019}. The empty grey `star' symbols designate the potential Sgr population from \citet{Hayes2020}. The two concentric circles indicate 5 $r_{t}$ and 7 $r_{t}$ for reference. Panel (b): \textit{Gaia} EDR3 proper motions of stars that have been associated with the Sgr stream: blue symbols for the  \citet{Antoja2020} stars and open black `star' symbols for \citet{Hayes2020} stars. The orbital path of Sgr is shown by the dotted (backward) and solid (forward) purple line in panels (a) and (b), with the thick and thin lines showing the central orbit, and one hundred ensemble of orbits that shows the more probable regions of the space, which are crossed more frequently by the simulated orbit, respectively. Panel (c): Color magnitude diagram from \textit{Gaia} EDR3 photometry of our sample. The symbols are the same as in panels (a) and (b), except the white circles, which denotes the M~54 members from \textit{Gaia} EDR3, selected on proper motions and within 3$\arcmin$ from the cluster center. Panel (d): Radial velocities versus [Fe/H] ratios determined from APOGEE-2/\texttt{ASPCAP} (black symbols) and our [Fe/H] ratio determinations from \texttt{BACCHUS} (green and red symbols) in the field around M~54. The [Fe/H] APOGEE-2/\texttt{ASPCAP} determinations have been systematically offset by $\sim$0.11 dex in order to compare with our [Fe/H] \texttt{BACCHUS}  determinations, as suggested in \citep{Fernandez-Trincado2020c}.}
		\label{Figure1}
	\end{center}
\end{figure}	
	
The Sagittarius (Sgr) dwarf spheroidal (dSph) galaxy is one of the closest massive satellites of the Milky Way (MW) \citep{Ibata1994}, and has yielded a wealth of observational evidence of ongoing accretion by the MW in the form of persistent stellar debris and tidal streams discovered by \citet{Mateo1996}, and extensively studied with photometric and spectroscopic observations over a huge range of distances ($\sim$10--100 kpc) \citep[see, e.g.,][]{Ibata2001, deBoer2015} using different stellar tracers--including Carbon stars \citep{Totten1998}, the first all-sky map of the tails using 2MASS M-giants \citep{Majewski2003}, red clump Stars \citep{Correnti2010}, RR Lyrae stars \citep{Newberg2003, Ramos2020}, and CN-strong stars \citep{Hanke2020}, among other tracers, usually in small patches along the stream (see, e.g., \citealt{Li2019}). These studies have been followed-up by numerical studies \citep[see, e.g.,][]{Law2005, Vasiliev2020}, as well as by using precise astrometry from the \textit{Gaia} second data release \citep[Gaia DR2;][]{Brown2018}, based on proper motions alone \citep{Antoja2020}. Its proximity provides a unique laboratory to study accretion in detail, through the tidally stripped streams that outflow from the Sgr system \citep[][]{Hasselquist2017, Hasselquist2019, Hayes2020}.
		
As a natural result of such an accretion event, there is a claim in the literature that not only field stars but also GCs have been accreted \citep[see, e.g.,][]{Massari2019}.  Some have been speculated to be lost in the disruption process, and may lie immersed in the Sgr stream.  Candidates include: M~54, Terzan 7, Arp 2, Terzan 8, Pal 12, Whiting 1, NGC 2419, NGC 6534, and NGC 4147 \citep[e.g.,][]{Law2010a, Bellazzini2020}, but a firm connection is still under debate \citep[e.g.,][]{Villanova2016, Tang2018, Huang2020, Yuan2020}. In this context, ``chemical tagging" \citep[e.g.,][]{Freeman2002}, which is based on the principle that the photospheric chemical compositions of stars reflect the site of their formation, is a promising route for investigation of this question.   

While the abundances of light and heavy elements for individual stars in GCs have been widely explored \citep[e.g.,][]{Pancino2017, Meszaros2020}, little is known about these abundances in disrupted GCs likely associated with the closest dwarf galaxies, such as Sgr \citep{Karlsson2012}. Although some evidence for chemical anomalies has been detected towards the inner bulge and halo of the MW \citep[see, e.g.,][]{Fernandez-Trincado2016b, Recio-Blanco2017, Schiavon2017, Fernandez-Trincado2017} and Local Group dwarf galaxies \citep[see, e.g.,][]{Fernandez-Trincado2020b}, suggesting the presence of GCs in the form of disrupted remnants, alternative ways to produce these stars have been recently discussed \citep{Bekki2019}.

This paper is outlined as follows. The high-resolution spectroscopic observations are discussed in Section \ref{section2}. Section \ref{section3} describes the sample associated with M~54, including a comparison with data from the literature. Section \ref{section4} presents our estimated stellar parameters and derived chemical-abundance determinations. Section \ref{section5} discusses the results, and our concluding remarks are presented in Section \ref{section6}.
  
\section{Data}
\label{section2}

  We make use of the internal dataset (which includes all data taken through March 2020) of the second-generation Apache Point Observatory Galactic Evolution Experiment \citep[APOGEE-2;][]{Majewski2017}, which includes the first observations from the Ir\'en\'ee du Pont 2.5-m Telescope at Las Campanas Observatory \citep[APO-2S;][]{Bowen1973} in the Southern Hemisphere (Chile), and more observations from the Sloan 2.5-m Telescope at Apache Point Observatory \citep[APO-2N;]{Gunn2006} in the Northern Hemisphere (New Mexico). The survey operates with two nearly identical spectrographs \citep{Eisenstein2011, Wilson2012, Wilson2019}, collecting high-resolution ($R\sim22,000$) spectra in the near-infrared textit{H}-band (1.5145--1.6960 $\mu$m, vacuum wavelengths). This data set provides stellar parameters, chemical abundances, and radial velocity (RV) information for more than 600,000 sources, which include $\sim$437,000 targets from the sixteenth data release \citep[DR16;][]{Ahumada2020} of the fourth generation of the Sloan Digital Sky Survey \citep[SDSS-IV;][]{Blanton2017}. APOGEE-2 target selection is described in full detail in \citet{Zasowski2017} (APOGEE-2), Santana et al. (in prep.) (APO-2S), and Beaton et al. (in prep.) (APO-2N). 
  
  APOGEE-2 spectra were reduced \citep{Nidever2015} and analyzed using the APOGEE Stellar Parameters and Chemical Abundance Pipeline \citep[ASPCAP;][]{Garcia2016, Holtzman2015, Holtzman2018, Henrik2018, Henrik2020}. The model grids for APOGEE-2 internal dataset are based on a complete set of \texttt{MARCS} stellar atmospheres \citep{Gustafsson2008}, which now extend to effective temperatures as low as 3200 K, and spectral synthesis using the \texttt{Turbospectrum} code \citep{Plez2012}. The APOGEE-2 spectra provide access to more than 26 chemical species, which are described in \citet{Smith2013}, \citet{Shetrone2015}, \citet{Hasselquist2016}, \citet{Cunha2017}, and \citet{Holtzman2018}.  

\subsection{M~54 field}
\label{section3}

The APOGEE-2 field toward M~54 was previously examined in \citet{Meszaros2020} based on public DR16 spectra. In that work, 22 stars were identified as potential members linked to M~54 based in the APOGEE-2 radial velocities \citep{Nidever2015}, i.e., stars with RV within $3\sigma _{RV,cluster}$, metallicity within $\pm$0.5 dex around the cluster average, proper motion from the \textit{Gaia} Early Data Release 3 \citep[\textit{Gaia} EDR3;][]{Brown2020} within 2.5$\sigma$ around the cluster average proper motion, and located inside the cluster tidal radius, $r_{t}\lesssim10$ arcmin, \citep[][2010 edition]{Harris1996} were classified as potential members of M~54. However, only 7 out of 22 stars were spectroscopically examined with the \texttt{BACCHUS} code in \citet{Masseron2016}, since only these stars achieved a signal-to-noise (S/N$>$70) sufficient to provide reliable abundance determinations.  

The post-APOGEE DR16 dataset provides incremental visits toward M~54, which has allowed to increase the signal-to-noise for 20 out of 22 of the potential cluster members. As a result, nitrogen, titanium, and nickel abundances can be now obtained from the stronger absorption features (as shown for $^{12}$C$^{14}$N lines as shown in Figure \ref{Figure4}), and other chemical species can also be studied. 

\citet{Nataf2019}, using APOGEE-2 DR14 data \citep{Abolfathi2018} and abundance determinations from the \texttt{Payne} pipeline \citep{Ting2019}, have catalogued eight possible members from M~54. Two of those objects (2M18544275$-$3029012 and 2M18550740$-$3026052) were included in our study. The remaining six stars were rejected from our analysis for the following reasons. Six objects in \citet{Nataf2019} were found to have low S/N ($<$70) spectra, resulting in very uncertain CNO abundance ratios for many chemical species, since the molecular lines ($^{16}$OH, $^{12}$C$^{16}$O, and $^{12}$C$^{14}$N) are very weak. Secondly, 6 out of the 8 objects in \citet{Nataf2019} exhibit [Fe/H] $> -1.1$, and were recently classified as Sgr stars \citep[see, e.g.,][]{Hayes2020}, which make them unlikely members of M~54.

In this study, we make use of the more recent spectra to examine the chemical composition of added stars to the abundance average of M~54. As in \citet{Meszaros2020}, we also limit our discussion only to stars with S/N$>70$.

\subsection{Extra-tidal stars}

We also report on the serendipitous discovery of two nitrogen-enhanced (N-rich) metal-poor stars beyond the tidal radius of M~54, as shown in pane (a) of Figure \ref{Figure1}. APOGEE-2 stars in the stream$+$core Sagittarius (Sgr) system \citep[see, e.g.,][]{Hasselquist2017, Hasselquist2019, Hayes2020} are highlighted as black open `star' symbols in pane (a) of Figure \ref{Figure1}, while potential star members (blue symbols) of the stream$+$core Sgr system from \citet[][]{Antoja2020} are also displayed in panel (a) of Figure \ref{Figure1}. It is important to note that the [Fe/H] abundance of APOGEE-2 Sgr stars are provided by the \texttt{ASPCAP} pipeline \citep[see][]{Hasselquist2017, Hasselquist2019, Hayes2020}. In order to compare with our [Fe/H] determinations, an offset of $\sim$0.11 dex was applied to \texttt{ASPCAP} metallicities in panel (d) of Figure \ref{Figure1}, as suggested in \citet{Fernandez-Trincado2020c}.

Panels (a) to (d) of Figure \ref{Figure1} reveal that one (2M18565969$-$3106454) of the newly discovered N-rich stars meets the minimum criterion to be considered a potential extra-tidal star which has likely escaped the cluster potential, while the second N-rich star (2M18533777$-$3129187) has physical properties that are clearly offset from the M~54 population. In particular, this star is brighter than the typical population of M~54 (see panel (c) in Figure \ref{Figure1}), and both proper motions and RV differ from the nominal proper motion and RV of the cluster as shown in panels (b) and (d) of Figures \ref{Figure1}. It is likely that 2M18533777$-$3129187 is a foreground field star (hereafter N-rich field star). 


\section{Stellar parameters and chemical-abundance determinations}
 \label{section4}

The chemical analysis is very similar to that carried out by \citet[][]{Fernandez-Trincado2019a, Fernandez-Trincado2019b, Fernandez-Trincado2019c, Fernandez-Trincado2019d, Fernandez-Trincado2020a, Fernandez-Trincado2020b, Fernandez-Trincado2020c, Fernandez-Trincado2020d, Fernandez-Trincado2021a}. The stellar parameters ($T_{\rm eff}$, $\log$ \textit{g}, and first guess on metallicity) for the 20 cluster members with S/N$>$70 were extracted from \citet{Meszaros2020}, while we adopt the atmospheric parameters from the uncalibrated post-APOGEE DR16 values for the two stars beyond the cluster tidal radius. The elemental abundances and final errors in [Fe/H] and [X/Fe], astrometric and kinematic properties of our sample are listed in Tables \ref{Table1},  \ref{Table11}, and \ref{Table2}, respectively. 

A consistent chemical-abundance analysis was then carried out with the \texttt{BACCHUS} code \citep{Masseron2016}, from which we obtained the metallicities from Fe I lines, and abundances for twelve other chemical species belonging to the light- (C, N), $\alpha$- (O, Mg, Si, Ca, and Ti), Fe-peak (Ni), odd-Z (Al, K) and \textit{s}-process (Ce, Nd) elements.

 \begin{figure}	
 	\begin{center}
 		\includegraphics[width=88mm]{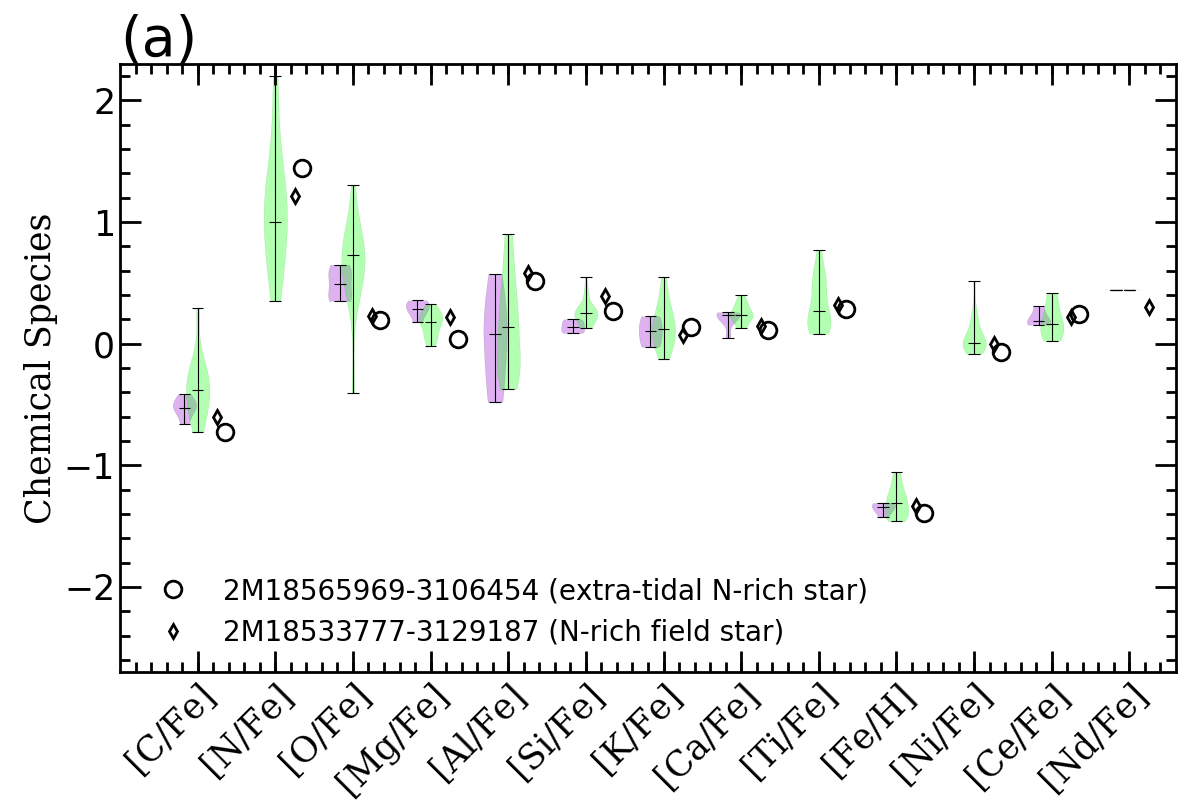}
 		\includegraphics[width=92mm]{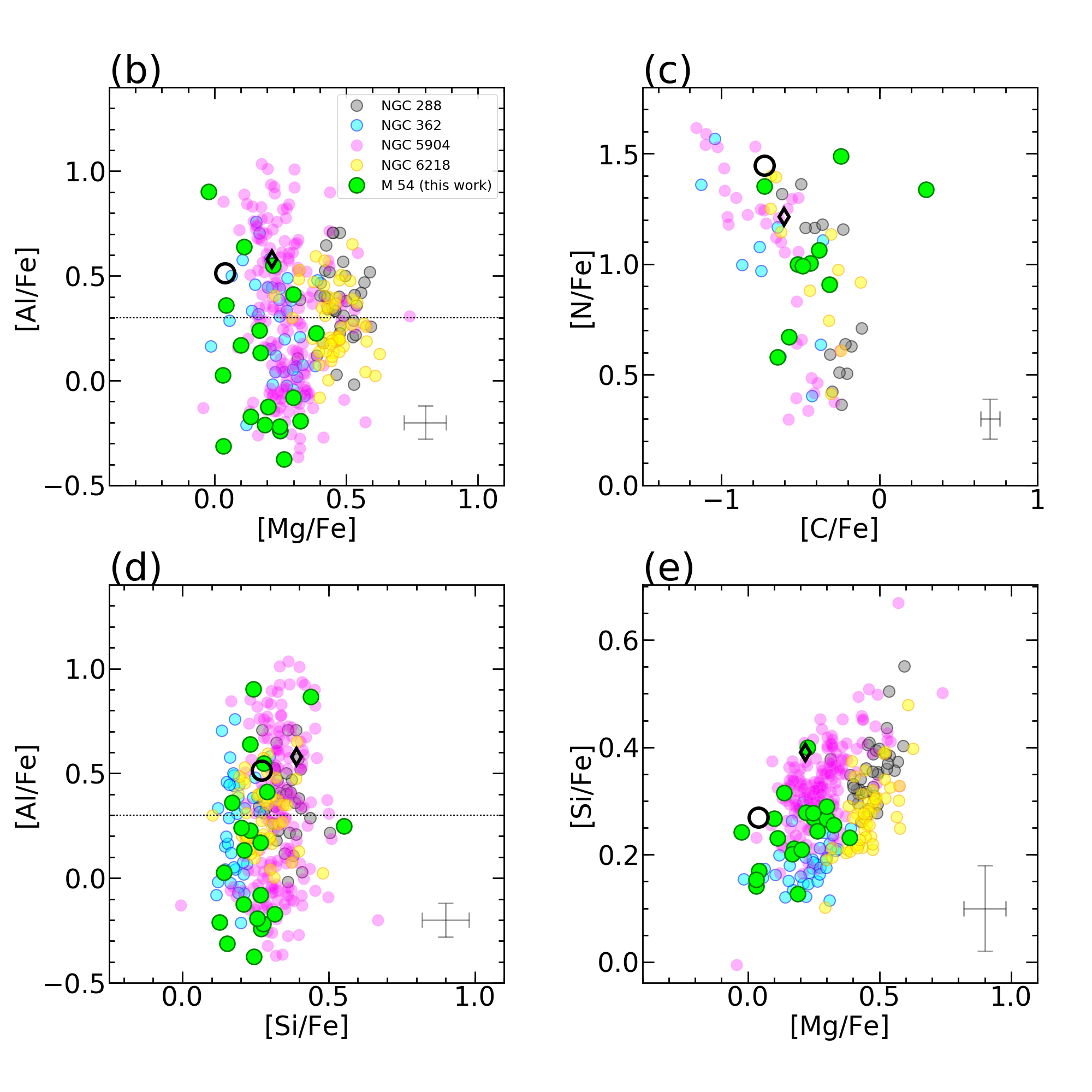}
 		\caption{{\bf \texttt{BACCHUS} elemental abundances}. Panel (a): The observed [X/H] and [Fe/H] abundance-density estimation (violin representation) of M~54 stars, and the observed abundance ratios of newly identified N-rich stars. The extra-tidal star from M~54 and a field star is highlighted with a black open circle and diamond, respectively. Each violin indicates with horizontal lines the median and limits of the distribution. The lime and dark violet violin representation refer to the abundance ratios of 20 stars (this work) and 7 stars from \citet{Meszaros2020}, respectively. Panels (b)--(e): Distributions of light- (C,N), $\alpha$- (Mg, Si) and odd-Z (Al) elements in different abundance planes. In each panel, the planes [Al/Fe] -- [Mg/Fe], [N/Fe]--[C/Fe], [Al/Fe]--[Si/Fe], [Si/Fe]--[Mg/Fe] are shown, respectively, for GCs from \citet{Meszaros2020}. The black dotted line at [Al/Fe] $=+0.3$ indicates the separation of FG and SG stars as proposed in \citet{Meszaros2020}. The distribution of M~54 stars (lime squares) analyzed in this work are overlaid. The black open circle and diamond refer to the extra-tidal and field N-rich star, respectively. The plotted error bars show the typical abundance uncertainties.}
 		\label{Figure2}
 	\end{center}
 \end{figure}	
 
   \section{Results and discussion}
 \label{section5}
 
 Panel (a) of Figure \ref{Figure2} summarizes the chemical enrichment seen in M~54 stars analyzed in this work, and compares to the \citet{Meszaros2020} determinations. The chemical composition of the two newly identified N-rich stars beyond the cluster tidal radius is also shown in the same figure. Overall, the chemical abundance of M~54 based on the added cluster stars is within the typical errors, and does not affect the science results presented in \citet{Meszaros2020}, while the two external N-rich stars share chemical patterns similar to the M~54 population.
 
 For M~54, we find a mean metallicity $\langle$[Fe/H]$\rangle = -1.30\pm0.12$, which agrees well with \citet{Meszaros2020}\footnote{Note that here, and for the abundances described below, the number following the average abundance represents the one-sigma dispersion, not the error in the mean.}. The spread in [Fe/H] increased from 0.04 to 0.12 dex, but it is still smaller than that reported in \citet{Carretta2010}. Even if the measured scatter is larger than that reported by \citet{Meszaros2020}, it does not seem to indicate the presence of a significant spread in [Fe/H], and is similar to that observed in Galactic globular clusters (GCs) at similar metallicity, such as M~10 \citep[see, e.g.,][]{Meszaros2020}. Nickel (an element that belongs to the Fe-group), exhibits a flat distribution as a function of [Fe/H], similar to that observed in \citet{Carretta2010}, and at odds with that observed in Sgr stars. 
 
Regarding the other chemical species, we find excellent agreement with the values provided by \citet{Meszaros2020}, as can be seen in panel (a) of Figure \ref{Figure2}, with the main difference that the added stars introduce a larger star-to-star scatter than previously measured. M~54 exhibits a modest enhancement in $\alpha$-elements, with mean values for [O/Fe], [Mg/Fe], [Si/Fe], [Ca/Fe], and [Ti/Fe] which is similar to what is seen in halo GCs: $\langle$[O/Fe]$\rangle = +0.64\pm0.36$ (14 stars); $\langle$[Mg/Fe]$\rangle = +0.18\pm0.11$ (18 stars); $\langle$[Si/Fe]$\rangle = +0.26\pm0.10$ (20 stars); $\langle$[Ca/Fe]$\rangle = +0.25\pm0.07$ (16 stars); and the new measured $\langle$[Ti/Fe]$\rangle = +0.21\pm0.21$ (16 stars), indicating a fast enrichment provided by supernovae (SNe) II events. Mean values are in good agreement with \citet{Meszaros2020}, with the exception of oxygen, which displays the larger star-to-star spread expected in likely second-generation stars. 

We also find that the [O/Fe], [Mg/Fe], and [Si/Fe] ratios are almost flat as a function of the metallicity, while [Ca/Fe] and [Ti/Fe] ratios slightly increases as [Fe/H] increases, similar to the behaviour found by \citet{Carretta2010}. On the contrary, the $\alpha$-element trend observed in Sgr stars \citep[see, e.g.,][]{Carretta2010, McWilliam2013, Hasselquist2017, Hasselquist2019} differ from those seen in the population of M~54. Overall, the $\alpha$-elements in the cluster are higher than seen in Sgr stars. In conclusion, the measured $\alpha$-enrichment in this work support the previous hypothesis suggesting that the $\alpha$-element in M~54 stars formed before the typical $e$-folding time for SN Ia contributing their ejecta to the gas pool \citep[e.g.,][]{Carretta2010}. 

We also found that some stars in M~54 appear to be quite Mg poor, with strong enrichment in aluminum and nitrogen, providing further evidence for the presence of second-generation stars in M~54, and the signature of very high temperatures achieved during H-burning \citep[e.g.,][]{Carretta2010, Meszaros2020}. The odd-Z elements (Al and K) in M~54 exhibit an average $\langle$[Al/Fe]$\rangle = +0.14\pm0.37$ (19 stars) and $\langle$[K/Fe]$\rangle = +0.15\pm0.18$ (17 stars), with a clear anti-correlation in Al-Mg, as can be seen in in panel (b) of Figure \ref{Figure2}, with moderate Mg depletions related to the enrichment in Al abundances, as the result of the conversion of Mg into Al during the Mg-Al cycle \citep[e.g.,][]{Carretta2010, Denissenkov2015, Renzini2015, Pancino2017}. This pattern is evidently not present in the Sgr stars, where, on the contrary, \texttt{ASPCAP} Mg and Al abundances are positively correlated with each other \citep[see, e.g.,][]{Hasselquist2017, Hasselquist2019, Hayes2020}. 

We derived average abundances for C and N in M~54, of $\langle$[C/Fe]$\rangle = -0.36\pm0.25$ (13 stars) and $\langle$[N/Fe]$\rangle = +1.12\pm0.48$ (17 stars). Most of the stars in M~54 are C deficient ([C/Fe]$\lesssim$+0.3) and N enhanced ([N/Fe]$>+0.5$), but they do not exhibit the typical N-C anti-correlation (see panel (c) of Figure \ref{Figure2}) seen in other GCs at similar metallicity \citep[e.g.][]{Meszaros2020}, most probably due the lack of stars with low nitrogen abundances. On the contrary, an apparent continuous distribution of N abundances is present in M~54. This result indicates the prevalence of the multiple-population phenomenon in M~54 as previously suggested in the literature \citep{Carretta2010, Milone2017, Sills2019, Meszaros2020}.

Additionally, we do not find any evidence for the presence of the K-Mg anti-correlation in M~54, as have been suggested to be present in a few Galactic GCs at similar metallicity \citep{Meszaros2020}. Furthermore, a Si-Al correlation is slightly evident in M~54, as shown in panel (d) of Figure \ref{Figure2}, and has a stubby Mg-Si distribution (see panel (e) of Figure \ref{Figure2}), which is an indication of $^{28}$Si production from the result of a secondary leakage in the main Mg-Al cycle, which is instead absent in the Sgr stars.
  
For the elements produced by neutron(\textit{n})-capture processes (Ce II and Nd II), we find, on average, $\langle$[Ce/Fe]$\rangle = +0.18\pm0.13$ (10 stars) and [Nd/Fe]$=+0.44$ (1 star). Overall, M~54 exhibits a modest enrichment in \textit{s}-process elements, with a few stars as enhanced as $+0.4$, similar to that observed in Galactic GC stars at similar metallicity \citep[see, e.g.,][]{Meszaros2020}, suggesting that it is possible that the \textit{s}-process enrichment has been produced by a different source than the progenitor of the Mg-Al anti-correlations, possibly by low-mass asymptotic giant branch stars. Lastly, we find that [Ce/Fe] ratios in M~54 are almost flat as a function of metrallicity. Unfortunately, Nd II is measured in only one star, which has been found to exhibit the modest enhancement, consistent with a moderate enrichment of \textit{s}-process elements.

Furthermore, we report the serendipitous discovery of two N-enhanced stars identified within $\sim$7$r_t$ from M~54, as shown in panel (a) of Figure \ref{Figure1}. Panel (a) of Figure \ref{Figure2} show the collection of [X/Fe] and [Fe/H] abundance ratios for the two newly identified N-rich stars beyond the tidal radius of M~54. Both stars exhibit very similar chemical-abundance patterns as those seen in the population of M~54. A plausible explanation is that both stars were previous members of M~54, from which they have been ejected. However, this possibility seems unlikely for one of these extra-tidal stars (2M18533777$-$3129187), which was ruled out as a possible member of M~54. 

As can be appreciated from inspection of panels (a) to (d) of Figure \ref{Figure1}, the current position of 2M18533777$-$3129187 does not resemble the kinematic and astrometric properties \citep[e.g.,][]{Antoja2020, Hayes2020} of Sgr+M~54 stars, nor the orbital path of Sgr\footnote{The Sgr orbit was computed with the \texttt{GravPot16} model, \url{https://gravpot.utinam.cnrs.fr}, by adopting the same model configurations as described in \citet{Fernandez-Trincado2020c}. For the Sgr centre, we adopt the heliocentric distance $d_{\odot} =$ 26.5 kpc and heliocentric radial velocity $RV = 142$ km s$^{-1}$ from \citet{Vasiliev2020b}, and proper motions from \citet{Helmi2018}: $\mu_{\alpha}\cos{\delta} = -2.692$ mas yr$^{-1}$ and $\mu_{\delta} = -1.359$ mas yr$^{-1}$, with uncertainties assumed of the order of 10\% in $d_{\odot}$, $RV$, and proper motions.} . It is also the most luminous star in our sample, making it a likely foreground star. The possibility that this star was disrupted from M~54 and deposited in the inner Galaxy seems unlikely, as the perigalacticon of M~54 is located well beyond the solar radius \citep[see, e.g.,][]{Baumgardt2019}. We   conclude that 2M18533777$-$3129187 is a N-enhanced field star born in a different progenitor than M~54, but with a similar chemical-enrichment history to this cluster.

 Aside from 2M18533777$-$3129187, there is another N-enhanced field star (2M18565969$-$3106454) located $\sim{}5\times{}r_{t}$ from the cluster center, which exhibits a stellar atmosphere strongly enriched in nitrogen ([N/Fe]$>+1.4$), as extreme as M~54 stars, accompanied by a very low carbon abundance ([C/Fe]$<-0.7$), and with discernible contributions from the \textit{s}-process elements (Ce II). Since the [Al/Fe] ratio is $>+0.5$, which is a ‘typical’ value for stars in GCs, and unlikely in dwarf galaxy populations, we conclude that 2M18565969$-$3106454 shares the same nucleosynthetic pathways of second-generation stars in M~54. 
 
2M18565969$-$3106454 is a potential extra-tidal star with kinematics and astrometric properties similar to that of M~54 stars, and exhibits unique chemical patterns comparable to that of genuine second-generation GC stars, which makes it very different from Sgr stars. On the other hand, N-rich stars are commonly observed to be more centrally concentrated in GCs \citep[e.g.][]{Dalessandro2019} and as a consequence they have smaller probabilities to be tidally stripped. Thus, it is likely that the extra-tidal star could well be just a stripped M~54 star as many others in its surroundings. Our finding demonstrate that N-rich stars are a promising route for identifying the unambiguous chemical signatures of stars formed in GC-like environment which may lie immersed in the M~54+Sgr core and/or Sgr stream, as well as confirm or discard the possible association of GCs to the Sgr stream \citep{Bellazzini2020}. 
 
Following the same methodology as described in \citet{Fernandez-Trincado2021b}, we compute the predicted number ($N_{N-rich}$) of N-rich field stars observed in APOGEE-2 toward M~54/Sgr using the smooth halo density relations presented in \citet{Horta2021}, and by adopting the same Monte Carlo implementation of the Von Neumann Rejection Technique \citep[see e.g.,][]{Press2002} as in Eq. 7 in \citep{Fernandez-Trincado2015a}. We find the expected number of observed N-rich halo stars beyond $d_{\odot}\gtrsim15$ kpc over the sky area of 1.5 degree radius centred in M~54, and with both astrometric and kinematic properties as M~54 to be $N_{N-rich}< 0.1$ (from 1000 Monte Carlo realisations). This yield a very low probability that the new identified extra-tidal N-rich star associated with M~54 is due to random fluctuations in the field. Furthermore, we also use the Besan\c{}con galactic model \citep{Robin2003} and the \texttt{GravPot16} model \citep{Fernandez-Trincado2020e} to explore the expectations for a "default" Milky Way along the RVs to the Sgr+M\~54 surrounding field beyond  $d_{\odot}\gtrsim$ 15 kpc. The "all" sample is dominated by halo kinematics with a negligible contribution from the thin and thick disk beyond $RV \gtrsim 120$ km s$^{-1}$.  Thus, our Milky Way simulated sample act to guide us in $RV$ space, confirming that the kinematics of the newly identified extra-tidal N-rich star differs from the disk population, with practically low contribution form the expected halo. , 

 \section{Concluding remarks}
 \label{section6}

We present a spectroscopic analysis for 20 out 22 red giant stars that are members of M~54 from the internal APOGEE DR16 dataset. This study doubles the sample of stars with spectroscopic measurements for this cluster, and the new post-APOGEE DR16 spectra achieve high signal-to-noise (S/N$>70$), allowing the addition of new chemical species not examined in previous studies \citep[e.g.,][]{Meszaros2020} in the \textit{H}-band--APOGEE-2 footprint. 
 
 Overall, the chemical species re-examined in M~54 were found to be consistent with previous studies \citep{Meszaros2020}, although most of them exhibit a large star-to-star scatter. We find that 15 out of the 20 stars investigated show a high [N/Fe] abundance ratio ([N/Fe]$\gtrsim+0.5$), confirming the prevalence of the MPs phenomenon in M~54. Both [Ni/Fe] and [Ti/Fe], not previously examined in \citet{Meszaros2020}, were found to be in good agreement with measurements in the literature. In particular, we confirm the [Ti/Fe]~ratio slightly increases as [Fe/H] increases, as has been reported in \citet{Carretta2010}. We also find a large spread in [Al/Fe], and the presence of a genuine second-generation star in M~54, which exhibits Mg deficiency ([Mg/Fe]$<$0) accompanied with large enhancements in nitrogen and aluminum. In general, all chemical species examined in the M~54 members present distinguishable chemical behaviour compared with Sgr stars, suggesting a different chemical-evolution history that resembles other Galactic halo GCs at similar metallicity.

Furthermore, we report on the serendipitous discovery of a potential extra-tidal star toward the surrounding regions of the M~54+Sgr core, which exhibits a strong enrichment in nitrogen comparable to that seen in M~54 stars. As far as we know this is the first study reporting on the unambiguous chemical signatures of stars formed in GC-like environment into a nearby satellite dwarf galaxy around the Milky Way. Finding out how many of such chemical unusual stars likely originated in GCs are present in dwarf galaxy systems, help to understand the link between GCs and their stellar streams \citep[see e.g.,][]{Bellazzini2020}. 

	\begin{acknowledgements}  
	The author is grateful for the enlightening feedback from the anonymous referee.
	 J.G.F-T is supported by FONDECYT No. 3180210. 
	 T.C.B. acknowledges partial support for this work from grant PHY 14-30152: Physics Frontier Center / JINA Center for the Evolution of the Elements (JINA-CEE), awarded by the US National Science Foundation. 
	 D.M. is supported by the BASAL Center for Astrophysics and Associated Technologies (CATA) through grant AFB 170002, and by project FONDECYT Regular No. 1170121. 
	 S.V. gratefully acknowledges the support provided by Fondecyt regular No. 1170518. 
    D.G. gratefully acknowledges support from the Chilean Centro de Excelencia en Astrof\'isica y Tecnolog\'ias Afines (CATA) BASAL grant AFB-170002. D.G. also acknowledges financial support from the Direcci\'on de Investigaci\'on y Desarrollo de la Universidad de La Serena through the Programa de Incentivo a la Investigaci\'on de Acad\'emicos (PIA-DIDULS). 
	 A.R.-L. acknowledges financial support provided in Chile by Agencia Nacional de Investigaci\'on y Desarrollo (ANID) through the FONDECYT project 1170476.
	 B.B. acknowledge partial financial support from the Brazilian agencies CAPES-Financial code 001, CNPq, and FAPESP. 
	 \\
	
	This work has made use of data from the European Space Agency (ESA) mission Gaia (\url{http://www.cosmos.esa.int/gaia}), processed by the Gaia Data Processing and Analysis Consortium (DPAC, \url{http://www.cosmos.esa.int/web/gaia/dpac/consortium}). Funding for the DPAC has been provided by national institutions, in particular the institutions participating in the Gaia Multilateral Agreement.\\
	
	Funding for the Sloan Digital Sky Survey IV has been provided by the Alfred P. Sloan Foundation, the U.S. Department of Energy Office of Science, and the Participating Institutions. SDSS- IV acknowledges support and resources from the Center for High-Performance Computing at the University of Utah. The SDSS web site is www.sdss.org. SDSS-IV is managed by the Astrophysical Research Consortium for the Participating Institutions of the SDSS Collaboration including the Brazilian Participation Group, the Carnegie Institution for Science, Carnegie Mellon University, the Chilean Participation Group, the French Participation Group, Harvard-Smithsonian Center for Astrophysics, Instituto de Astrof\`{i}sica de Canarias, The Johns Hopkins University, Kavli Institute for the Physics and Mathematics of the Universe (IPMU) / University of Tokyo, Lawrence Berkeley National Laboratory, Leibniz Institut f\"{u}r Astrophysik Potsdam (AIP), Max-Planck-Institut f\"{u}r Astronomie (MPIA Heidelberg), Max-Planck-Institut f\"{u}r Astrophysik (MPA Garching), Max-Planck-Institut f\"{u}r Extraterrestrische Physik (MPE), National Astronomical Observatory of China, New Mexico State University, New York University, University of Notre Dame, Observat\'{o}rio Nacional / MCTI, The Ohio State University, Pennsylvania State University, Shanghai Astronomical Observatory, United Kingdom Participation Group, Universidad Nacional Aut\'{o}noma de M\'{e}xico, University of Arizona, University of Colorado Boulder, University of Oxford, University of Portsmouth, University of Utah, University of Virginia, University of Washington, University of Wisconsin, Vanderbilt University, and Yale University.\\
\end{acknowledgements}
	
\bibliographystyle{aa}
\bibliography{references}

\begin{thebibliography}{86}
\expandafter\ifx\csname natexlab\endcsname\relax\def\natexlab#1{#1}\fi

\bibitem[{{Abolfathi} {et~al.}(2018){Abolfathi}, {Aguado}, {Aguilar}, {Allende
  Prieto}, {Almeida}, {Ananna}, {Anders}, {Anderson}, {Andrews}, {Anguiano},
  {Arag{\'o}n-Salamanca}, {Argudo-Fern{\'a}ndez}, {Armengaud}, {Ata},
  {Aubourg}, {Avila-Reese}, {Badenes}, {Bailey}, {Balland}, {Barger},
  {Barrera-Ballesteros}, {Bartosz}, {Bastien}, {Bates}, {Baumgarten},
  {Bautista}, {Beaton}, {Beers}, {Belfiore}, {Bender}, {Bernardi}, {Bershady},
  {Beutler}, {Bird}, {Bizyaev}, {Blanc}, {Blanton}, {Blomqvist}, {Bolton},
  {Boquien}, {Borissova}, {Bovy}, {Bradna Diaz}, {Brandt}, {Brinkmann},
  {Brownstein}, {Bundy}, {Burgasser}, {Burtin}, {Busca}, {Ca{\~n}as},
  {Cano-D{\'\i}az}, {Cappellari}, {Carrera}, {Casey}, {Cervantes Sodi}, {Chen},
  {Cherinka}, {Chiappini}, {Choi}, {Chojnowski}, {Chuang}, {Chung}, {Clerc},
  {Cohen}, {Comerford}, {Comparat}, {Correa do Nascimento}, {da Costa},
  {Cousinou}, {Covey}, {Crane}, {Cruz-Gonzalez}, {Cunha}, {da Silva Ilha},
  {Damke}, {Darling}, {Davidson}, {Dawson}, {de Icaza Lizaola}, {de la
  Macorra}, {de la Torre}, {De Lee}, {de Sainte Agathe}, {Deconto Machado},
  {Dell'Agli}, {Delubac}, {Diamond-Stanic}, {Donor}, {Downes}, {Drory}, {du Mas
  des Bourboux}, {Duckworth}, {Dwelly}, {Dyer}, {Ebelke}, {Davis Eigenbrot},
  {Eisenstein}, {Elsworth}, {Emsellem}, {Eracleous}, {Erfanianfar},
  {Escoffier}, {Fan}, {Fern{\'a}ndez Alvar}, {Fernandez-Trincado}, {Fernand o
  Cirolini}, {Feuillet}, {Finoguenov}, {Fleming}, {Font-Ribera}, {Freischlad},
  {Frinchaboy}, {Fu}, {G{\'o}mez Maqueo Chew}, {Galbany}, {Garc{\'\i}a
  P{\'e}rez}, {Garcia-Dias}, {Garc{\'\i}a-Hern{\'a}ndez}, {Garma Oehmichen},
  {Gaulme}, {Gelfand }, {Gil-Mar{\'\i}n}, {Gillespie}, {Goddard}, {Gonz{\'a}lez
  Hern{\'a}ndez}, {Gonzalez-Perez}, {Grabowski}, {Green}, {Grier}, {Gueguen},
  {Guo}, {Guy}, {Hagen}, {Hall}, {Harding}, {Hasselquist}, {Hawley}, {Hayes},
  {Hearty}, {Hekker}, {Hernand ez}, {Hernandez Toledo}, {Hogg},
  {Holley-Bockelmann}, {Holtzman}, {Hou}, {Hsieh}, {Hunt}, {Hutchinson},
  {Hwang}, {Jimenez Angel}, {Johnson}, {Jones}, {J{\"o}nsson}, {Jullo}, {Khan},
  {Kinemuchi}, {Kirkby}, {Kirkpatrick}, {Kitaura}, {Knapp}, {Kneib},
  {Kollmeier}, {Lacerna}, {Lane}, {Lang}, {Law}, {Le Goff}, {Lee}, {Li}, {Li},
  {Lian}, {Liang}, {Lima}, {Lin}, {Long}, {Lucatello}, {Lundgren}, {Mackereth},
  {MacLeod}, {Mahadevan}, {Maia}, {Majewski}, {Manchado}, {Maraston},
  {Mariappan}, {Marques-Chaves}, {Masseron}, {Masters}, {McDermid}, {McGreer},
  {Melendez}, {Meneses-Goytia}, {Merloni}, {Merrifield}, {Meszaros}, {Meza},
  {Minchev}, {Minniti}, {Mueller}, {Muller-Sanchez}, {Muna}, {Mu{\~n}oz},
  {Myers}, {Nair}, {Nand ra}, {Ness}, {Newman}, {Nichol}, {Nidever},
  {Nitschelm}, {Noterdaeme}, {O'Connell}, {Oelkers}, {Oravetz}, {Oravetz},
  {Ort{\'\i}z}, {Osorio}, {Pace}, {Padilla}, {Palanque-Delabrouille},
  {Palicio}, {Pan}, {Pan}, {Parikh}, {P{\^a}ris}, {Park}, {Peirani},
  {Pellejero-Ibanez}, {Penny}, {Percival}, {Perez-Fournon}, {Petitjean},
  {Pieri}, {Pinsonneault}, {Pisani}, {Prada}, {Prakash}, {Queiroz}, {Raddick},
  {Raichoor}, {Barboza Rembold}, {Richstein}, {Riffel}, {Riffel}, {Rix},
  {Robin}, {Rodr{\'\i}guez Torres}, {Rom{\'a}n-Z{\'u}{\~n}iga}, {Ross},
  {Rossi}, {Ruan}, {Ruggeri}, {Ruiz}, {Salvato}, {S{\'a}nchez}, {S{\'a}nchez},
  {Sanchez Almeida}, {S{\'a}nchez-Gallego}, {Santana Rojas}, {Santiago},
  {Schiavon}, {Schimoia}, {Schlafly}, {Schlegel}, {Schneider}, {Schuster},
  {Schwope}, {Seo}, {Serenelli}, {Shen}, {Shen}, {Shetrone}, {Shull}, {Silva
  Aguirre}, {Simon}, {Skrutskie}, {Slosar}, {Smethurst}, {Smith}, {Sobeck},
  {Somers}, {Souter}, {Souto}, {Spindler}, {Stark}, {Stassun}, {Steinmetz},
  {Stello}, {Storchi-Bergmann}, {Streblyanska}, {Stringfellow}, {Su{\'a}rez},
  {Sun}, {Szigeti}, {Taghizadeh-Popp}, {Talbot}, {Tang}, {Tao}, {Tayar},
  {Tembe}, {Teske}, {Thakar}, {Thomas}, {Tissera}, {Tojeiro}, {Tremonti},
  {Troup}, {Urry}, {Valenzuela}, {van den Bosch}, {Vargas-Gonz{\'a}lez},
  {Vargas-Maga{\~n}a}, {Vazquez}, {Villanova}, {Vogt}, {Wake}, {Wang},
  {Weaver}, {Weijmans}, {Weinberg}, {Westfall}, {Whelan}, {Wilcots}, {Wild},
  {Williams}, {Wilson}, {Wood-Vasey}, {Wylezalek}, {Xiao}, {Yan}, {Yang},
  {Ybarra}, {Y{\`e}che}, {Zakamska}, {Zamora}, {Zarrouk}, {Zasowski}, {Zhang},
  {Zhao}, {Zhao}, {Zheng}, {Zheng}, {Zhou}, {Zhu}, {Zinn}, \&
  {Zou}}]{Abolfathi2018}
{Abolfathi}, B., {Aguado}, D.~S., {Aguilar}, G., {et~al.} 2018, \apjs, 235, 42

\bibitem[{{Ahumada} {et~al.}(2020){Ahumada}, {Allende Prieto}, {Almeida},
  {Anders}, {Anderson}, {Andrews}, {Anguiano}, {Arcodia}, {Armengaud},
  {Aubert}, {Avila}, {Avila-Reese}, {Badenes}, {Balland }, {Barger},
  {Barrera-Ballesteros}, {Basu}, {Bautista}, {Beaton}, {Beers}, {Benavides},
  {Bender}, {Bernardi}, {Bershady}, {Beutler}, {Bidin}, {Bird}, {Bizyaev},
  {Blanc}, {Blanton}, {Boquien}, {Borissova}, {Bovy}, {Brand t}, {Brinkmann},
  {Brownstein}, {Bundy}, {Bureau}, {Burgasser}, {Burtin}, {Cano-D{\'\i}az},
  {Capasso}, {Cappellari}, {Carrera}, {Chabanier}, {Chaplin}, {Chapman},
  {Cherinka}, {Chiappini}, {Doohyun Choi}, {Chojnowski}, {Chung}, {Clerc},
  {Coffey}, {Comerford}, {Comparat}, {da Costa}, {Cousinou}, {Covey}, {Crane},
  {Cunha}, {da Silva Ilha}, {Dai}, {Damsted}, {Darling}, {Davidson}, {Davies},
  {Dawson}, {De}, {de la Macorra}, {De Lee}, {de Andrade Queiroz}, {Deconto
  Machado}, {de la Torre}, {Dell'Agli}, {du Mas des Bourboux},
  {Diamond-Stanic}, {Dillon}, {Donor}, {Drory}, {Duckworth}, {Dwelly},
  {Ebelke}, {Eftekharzadeh}, {Eigenbrot}, {Elsworth}, {Eracleous},
  {Erfanianfar}, {Escoffier}, {Fan}, {Farr}, {Fern{\'a}ndez-Trincado},
  {Feuillet}, {Finoguenov}, {Fofie}, {Fraser-McKelvie}, {Frinchaboy},
  {Fromenteau}, {Fu}, {Galbany}, {Garcia}, {Garc{\'\i}a-Hern{\'a}ndez}, {Garma
  Oehmichen}, {Ge}, {Geimba Maia}, {Geisler}, {Gelfand }, {Goddy},
  {Gonzalez-Perez}, {Grabowski}, {Green}, {Grier}, {Guo}, {Guy}, {Harding},
  {Hasselquist}, {Hawken}, {Hayes}, {Hearty}, {Hekker}, {Hogg}, {Holtzman},
  {Horta}, {Hou}, {Hsieh}, {Huber}, {Hunt}, {Ider Chitham}, {Imig}, {Jaber},
  {Jimenez Angel}, {Johnson}, {Jones}, {J{\"o}nsson}, {Jullo}, {Kim},
  {Kinemuchi}, {Kirkpatrick}, {Kite}, {Klaene}, {Kneib}, {Kollmeier}, {Kong},
  {Kounkel}, {Krishnarao}, {Lacerna}, {Lan}, {Lane}, {Law}, {Le Goff}, {Leung},
  {Lewis}, {Li}, {Lian}, {Lin}, {Long}, {Longa-Pe{\~n}a}, {Lundgren}, {Lyke},
  {Ted Mackereth}, {MacLeod}, {Majewski}, {Manchado}, {Maraston}, {Martini},
  {Masseron}, {Masters}, {Mathur}, {McDermid}, {Merloni}, {Merrifield},
  {M{\'e}sz{\'a}ros}, {Miglio}, {Minniti}, {Minsley}, {Miyaji}, {Mohammad},
  {Mosser}, {Mueller}, {Muna}, {Mu{\~n}oz-Guti{\'e}rrez}, {Myers}, {Nadathur},
  {Nair}, {Nandra}, {do Nascimento}, {Nevin}, {Newman}, {Nidever}, {Nitschelm},
  {Noterdaeme}, {O'Connell}, {Olmstead}, {Oravetz}, {Oravetz}, {Osorio},
  {Pace}, {Padilla}, {Palanque-Delabrouille}, {Palicio}, {Pan}, {Pan},
  {Parker}, {Paviot}, {Peirani}, {Pe{\~n}a Ram{\'r}ez}, {Penny}, {Percival},
  {Perez-Fournon}, {P{\'e}rez-R{\`a}fols}, {Petitjean}, {Pieri},
  {Pinsonneault}, {Poovelil}, {Povick}, {Prakash}, {Price-Whelan}, {Raddick},
  {Raichoor}, {Ray}, {Rembold}, {Rezaie}, {Riffel}, {Riffel}, {Rix}, {Robin},
  {Roman-Lopes}, {Rom{\'a}n-Z{\'u}{\~n}iga}, {Rose}, {Ross}, {Rossi}, {Rowland
  s}, {Rubin}, {Salvato}, {S{\'a}nchez}, {S{\'a}nchez-Menguiano},
  {S{\'a}nchez-Gallego}, {Sayres}, {Schaefer}, {Schiavon}, {Schimoia},
  {Schlafly}, {Schlegel}, {Schneider}, {Schultheis}, {Schwope}, {Seo},
  {Serenelli}, {Shafieloo}, {Shamsi}, {Shao}, {Shen}, {Shetrone}, {Shirley},
  {Silva Aguirre}, {Simon}, {Skrutskie}, {Slosar}, {Smethurst}, {Sobeck},
  {Sodi}, {Souto}, {Stark}, {Stassun}, {Steinmetz}, {Stello}, {Stermer},
  {Storchi-Bergmann}, {Streblyanska}, {Stringfellow}, {Stutz}, {Su{\'a}rez},
  {Sun}, {Taghizadeh-Popp}, {Talbot}, {Tayar}, {Thakar}, {Theriault}, {Thomas},
  {Thomas}, {Tinker}, {Tojeiro}, {Toledo}, {Tremonti}, {Troup}, {Tuttle},
  {Unda-Sanzana}, {Valentini}, {Vargas-Gonz{\'a}lez}, {Vargas-Maga{\~n}a},
  {V{\'a}zquez-Mata}, {Vivek}, {Wake}, {Wang}, {Weaver}, {Weijmans}, {Wild},
  {Wilson}, {Wilson}, {Wolthuis}, {Wood-Vasey}, {Yan}, {Yang}, {Y{\`e}che},
  {Zamora}, {Zarrouk}, {Zasowski}, {Zhang}, {Zhao}, {Zhao}, {Zheng}, {Zheng},
  {Zhu}, \& {Zou}}]{Ahumada2020}
{Ahumada}, R., {Allende Prieto}, C., {Almeida}, A., {et~al.} 2020, \apjs, 249,
  3

\bibitem[{{Antoja} {et~al.}(2020){Antoja}, {Ramos}, {Mateu}, {Helmi}, {Anders},
  {Jordi}, \& {Carballo-Bello}}]{Antoja2020}
{Antoja}, T., {Ramos}, P., {Mateu}, C., {et~al.} 2020, \aap, 635, L3

\bibitem[{{Baumgardt} {et~al.}(2019){Baumgardt}, {Hilker}, {Sollima}, \&
  {Bellini}}]{Baumgardt2019}
{Baumgardt}, H., {Hilker}, M., {Sollima}, A., \& {Bellini}, A. 2019, \mnras,
  482, 5138

\bibitem[{{Bekki}(2019)}]{Bekki2019}
{Bekki}, K. 2019, \mnras, 490, 4007

\bibitem[{{Bellazzini} {et~al.}(2020){Bellazzini}, {Ibata}, {Malhan}, {Martin},
  {Famaey}, \& {Thomas}}]{Bellazzini2020}
{Bellazzini}, M., {Ibata}, R., {Malhan}, K., {et~al.} 2020, \aap, 636, A107

\bibitem[{{Blanton} {et~al.}(2017){Blanton}, {Bershady}, {Abolfathi},
  {Albareti}, {Allende Prieto}, {Almeida}, {Alonso-Garc{\'\i}a}, {Anders},
  {Anderson}, {Andrews}, {Aquino-Ort{\'\i}z}, {Arag{\'o}n-Salamanca},
  {Argudo-Fern{\'a}ndez}, {Armengaud}, {Aubourg}, {Avila-Reese}, {Badenes},
  {Bailey}, {Barger}, {Barrera-Ballesteros}, {Bartosz}, {Bates}, {Baumgarten},
  {Bautista}, {Beaton}, {Beers}, {Belfiore}, {Bender}, {Berlind}, {Bernardi},
  {Beutler}, {Bird}, {Bizyaev}, {Blanc}, {Blomqvist}, {Bolton}, {Boquien},
  {Borissova}, {van den Bosch}, {Bovy}, {Brandt}, {Brinkmann}, {Brownstein},
  {Bundy}, {Burgasser}, {Burtin}, {Busca}, {Cappellari}, {Delgado Carigi},
  {Carlberg}, {Carnero Rosell}, {Carrera}, {Chanover}, {Cherinka}, {Cheung},
  {G{\'o}mez Maqueo Chew}, {Chiappini}, {Choi}, {Chojnowski}, {Chuang},
  {Chung}, {Cirolini}, {Clerc}, {Cohen}, {Comparat}, {da Costa}, {Cousinou},
  {Covey}, {Crane}, {Croft}, {Cruz-Gonzalez}, {Garrido Cuadra}, {Cunha},
  {Damke}, {Darling}, {Davies}, {Dawson}, {de la Macorra}, {Dell'Agli}, {De
  Lee}, {Delubac}, {Di Mille}, {Diamond-Stanic}, {Cano-D{\'\i}az}, {Donor},
  {Downes}, {Drory}, {du Mas des Bourboux}, {Duckworth}, {Dwelly}, {Dyer},
  {Ebelke}, {Eigenbrot}, {Eisenstein}, {Emsellem}, {Eracleous}, {Escoffier},
  {Evans}, {Fan}, {Fern{\'a}ndez-Alvar}, {Fernandez-Trincado}, {Feuillet},
  {Finoguenov}, {Fleming}, {Font-Ribera}, {Fredrickson}, {Freischlad},
  {Frinchaboy}, {Fuentes}, {Galbany}, {Garcia-Dias},
  {Garc{\'\i}a-Hern{\'a}ndez}, {Gaulme}, {Geisler}, {Gelfand},
  {Gil-Mar{\'\i}n}, {Gillespie}, {Goddard}, {Gonzalez-Perez}, {Grabowski},
  {Green}, {Grier}, {Gunn}, {Guo}, {Guy}, {Hagen}, {Hahn}, {Hall}, {Harding},
  {Hasselquist}, {Hawley}, {Hearty}, {Gonzalez Hern{\'a}ndez}, {Ho}, {Hogg},
  {Holley-Bockelmann}, {Holtzman}, {Holzer}, {Huehnerhoff}, {Hutchinson},
  {Hwang}, {Ibarra-Medel}, {da Silva Ilha}, {Ivans}, {Ivory}, {Jackson},
  {Jensen}, {Johnson}, {Jones}, {J{\"o}nsson}, {Jullo}, {Kamble}, {Kinemuchi},
  {Kirkby}, {Kitaura}, {Klaene}, {Knapp}, {Kneib}, {Kollmeier}, {Lacerna},
  {Lane}, {Lang}, {Law}, {Lazarz}, {Lee}, {Le Goff}, {Liang}, {Li}, {Li},
  {Lian}, {Lima}, {Lin}, {Lin}, {Bertran de Lis}, {Liu}, {de Icaza Lizaola},
  {Long}, {Lucatello}, {Lundgren}, {MacDonald}, {Deconto Machado}, {MacLeod},
  {Mahadevan}, {Geimba Maia}, {Maiolino}, {Majewski}, {Malanushenko},
  {Malanushenko}, {Manchado}, {Mao}, {Maraston}, {Marques-Chaves}, {Masseron},
  {Masters}, {McBride}, {McDermid}, {McGrath}, {McGreer}, {Medina Pe{\~n}a},
  {Melendez}, {Merloni}, {Merrifield}, {Meszaros}, {Meza}, {Minchev},
  {Minniti}, {Miyaji}, {More}, {Mulchaey}, {M{\"u}ller-S{\'a}nchez}, {Muna},
  {Munoz}, {Myers}, {Nair}, {Nandra}, {Correa do Nascimento}, {Negrete},
  {Ness}, {Newman}, {Nichol}, {Nidever}, {Nitschelm}, {Ntelis}, {O'Connell},
  {Oelkers}, {Oravetz}, {Oravetz}, {Pace}, {Padilla}, {Palanque-Delabrouille},
  {Alonso Palicio}, {Pan}, {Parejko}, {Parikh}, {P{\^a}ris}, {Park}, {Patten},
  {Peirani}, {Pellejero-Ibanez}, {Penny}, {Percival}, {Perez-Fournon},
  {Petitjean}, {Pieri}, {Pinsonneault}, {Pisani}, {Poleski}, {Prada},
  {Prakash}, {Queiroz}, {Raddick}, {Raichoor}, {Barboza Rembold}, {Richstein},
  {Riffel}, {Riffel}, {Rix}, {Robin}, {Rockosi}, {Rodr{\'\i}guez-Torres},
  {Roman-Lopes}, {Rom{\'a}n-Z{\'u}{\~n}iga}, {Rosado}, {Ross}, {Rossi}, {Ruan},
  {Ruggeri}, {Rykoff}, {Salazar-Albornoz}, {Salvato}, {S{\'a}nchez}, {Aguado},
  {S{\'a}nchez-Gallego}, {Santana}, {Santiago}, {Sayres}, {Schiavon}, {da Silva
  Schimoia}, {Schlafly}, {Schlegel}, {Schneider}, {Schultheis}, {Schuster},
  {Schwope}, {Seo}, {Shao}, {Shen}, {Shetrone}, {Shull}, {Simon}, {Skinner},
  {Skrutskie}, {Slosar}, {Smith}, {Sobeck}, {Sobreira}, {Somers}, {Souto},
  {Stark}, {Stassun}, {Stauffer}, {Steinmetz}, {Storchi-Bergmann},
  {Streblyanska}, {Stringfellow}, {Su{\'a}rez}, {Sun}, {Suzuki}, {Szigeti},
  {Taghizadeh-Popp}, {Tang}, {Tao}, {Tayar}, {Tembe}, {Teske}, {Thakar},
  {Thomas}, {Thompson}, {Tinker}, {Tissera}, {Tojeiro}, {Hernandez Toledo}, {de
  la Torre}, {Tremonti}, {Troup}, {Valenzuela}, {Martinez Valpuesta},
  {Vargas-Gonz{\'a}lez}, {Vargas-Maga{\~n}a}, {Vazquez}, {Villanova}, {Vivek},
  {Vogt}, {Wake}, {Walterbos}, {Wang}, {Weaver}, {Weijmans}, {Weinberg},
  {Westfall}, {Whelan}, {Wild}, {Wilson}, {Wood-Vasey}, {Wylezalek}, {Xiao},
  {Yan}, {Yang}, {Ybarra}, {Y{\`e}che}, {Zakamska}, {Zamora}, {Zarrouk},
  {Zasowski}, {Zhang}, {Zhao}, {Zheng}, {Zheng}, {Zhou}, {Zhou}, {Zhu},
  {Zoccali}, \& {Zou}}]{Blanton2017}
{Blanton}, M.~R., {Bershady}, M.~A., {Abolfathi}, B., {et~al.} 2017, \aj, 154,
  28

\bibitem[{{Bowen} \& {Vaughan}(1973)}]{Bowen1973}
{Bowen}, I.~S. \& {Vaughan}, A.~H., J. 1973, \ao, 12, 1430

\bibitem[{{Carretta} {et~al.}(2010){Carretta}, {Bragaglia}, {Gratton},
  {Lucatello}, {Bellazzini}, {Catanzaro}, {Leone}, {Momany}, {Piotto}, \&
  {D'Orazi}}]{Carretta2010}
{Carretta}, E., {Bragaglia}, A., {Gratton}, R.~G., {et~al.} 2010, \aap, 520,
  A95

\bibitem[{{Correnti} {et~al.}(2010){Correnti}, {Bellazzini}, {Ibata},
  {Ferraro}, \& {Varghese}}]{Correnti2010}
{Correnti}, M., {Bellazzini}, M., {Ibata}, R.~A., {Ferraro}, F.~R., \&
  {Varghese}, A. 2010, \apj, 721, 329

\bibitem[{{Cunha} {et~al.}(2017){Cunha}, {Smith}, {Hasselquist}, {Souto},
  {Shetrone}, {Allende Prieto}, {Bizyaev}, {Frinchaboy},
  {Garc{\'\i}a-Hern{\'a}ndez}, {Holtzman}, {Johnson}, {J{\H{o}}nsson},
  {Majewski}, {M{\'e}sz{\'a}ros}, {Nidever}, {Pinsonneault}, {Schiavon},
  {Sobeck}, {Skrutskie}, {Zamora}, {Zasowski}, \&
  {Fern{\'a}ndez-Trincado}}]{Cunha2017}
{Cunha}, K., {Smith}, V.~V., {Hasselquist}, S., {et~al.} 2017, \apj, 844, 145

\bibitem[{{Dalessandro} {et~al.}(2019){Dalessandro}, {Cadelano}, {Vesperini},
  {Martocchia}, {Ferraro}, {Lanzoni}, {Bastian}, {Hong}, \&
  {Sanna}}]{Dalessandro2019}
{Dalessandro}, E., {Cadelano}, M., {Vesperini}, E., {et~al.} 2019, \apjl, 884,
  L24

\bibitem[{{de Boer} {et~al.}(2015){de Boer}, {Belokurov}, \&
  {Koposov}}]{deBoer2015}
{de Boer}, T.~J.~L., {Belokurov}, V., \& {Koposov}, S. 2015, \mnras, 451, 3489

\bibitem[{{Denissenkov} {et~al.}(2015){Denissenkov}, {VandenBerg}, {Hartwick},
  {Herwig}, {Weiss}, \& {Paxton}}]{Denissenkov2015}
{Denissenkov}, P.~A., {VandenBerg}, D.~A., {Hartwick}, F.~D.~A., {et~al.} 2015,
  \mnras, 448, 3314

\bibitem[{{Eisenstein} {et~al.}(2011){Eisenstein}, {Weinberg}, {Agol},
  {Aihara}, {Allende Prieto}, {Anderson}, {Arns}, {Aubourg}, {Bailey},
  {Balbinot}, {Barkhouser}, {Beers}, {Berlind}, {Bickerton}, {Bizyaev},
  {Blanton}, {Bochanski}, {Bolton}, {Bosman}, {Bovy}, {Brandt}, {Breslauer},
  {Brewington}, {Brinkmann}, {Brown}, {Brownstein}, {Burger}, {Busca},
  {Campbell}, {Cargile}, {Carithers}, {Carlberg}, {Carr}, {Chang}, {Chen},
  {Chiappini}, {Comparat}, {Connolly}, {Cortes}, {Croft}, {Cunha}, {da Costa},
  {Davenport}, {Dawson}, {De Lee}, {Porto de Mello}, {de Simoni}, {Dean},
  {Dhital}, {Ealet}, {Ebelke}, {Edmondson}, {Eiting}, {Escoffier}, {Esposito},
  {Evans}, {Fan}, {Femen{\'\i}a Castell{\'a}}, {Dutra Ferreira}, {Fitzgerald},
  {Fleming}, {Font-Ribera}, {Ford}, {Frinchaboy}, {Garc{\'\i}a P{\'e}rez},
  {Gaudi}, {Ge}, {Ghezzi}, {Gillespie}, {Gilmore}, {Girardi}, {Gott}, {Gould},
  {Grebel}, {Gunn}, {Hamilton}, {Harding}, {Harris}, {Hawley}, {Hearty},
  {Hennawi}, {Gonz{\'a}lez Hern{\'a}ndez}, {Ho}, {Hogg}, {Holtzman},
  {Honscheid}, {Inada}, {Ivans}, {Jiang}, {Jiang}, {Johnson}, {Jordan},
  {Jordan}, {Kauffmann}, {Kazin}, {Kirkby}, {Klaene}, {Knapp}, {Kneib},
  {Kochanek}, {Koesterke}, {Kollmeier}, {Kron}, {Lampeitl}, {Lang}, {Lawler},
  {Le Goff}, {Lee}, {Lee}, {Leisenring}, {Lin}, {Liu}, {Long}, {Loomis},
  {Lucatello}, {Lundgren}, {Lupton}, {Ma}, {Ma}, {MacDonald}, {Mack},
  {Mahadevan}, {Maia}, {Majewski}, {Makler}, {Malanushenko}, {Malanushenko},
  {Mand elbaum}, {Maraston}, {Margala}, {Maseman}, {Masters}, {McBride},
  {McDonald}, {McGreer}, {McMahon}, {Mena Requejo}, {M{\'e}nard},
  {Miralda-Escud{\'e}}, {Morrison}, {Mullally}, {Muna}, {Murayama}, {Myers},
  {Naugle}, {Neto}, {Nguyen}, {Nichol}, {Nidever}, {O'Connell}, {Ogando},
  {Olmstead}, {Oravetz}, {Padmanabhan}, {Paegert}, {Palanque-Delabrouille},
  {Pan}, {Pandey}, {Parejko}, {P{\^a}ris}, {Pellegrini}, {Pepper}, {Percival},
  {Petitjean}, {Pfaffenberger}, {Pforr}, {Phleps}, {Pichon}, {Pieri}, {Prada},
  {Price-Whelan}, {Raddick}, {Ramos}, {Reid}, {Reyle}, {Rich}, {Richards},
  {Rieke}, {Rieke}, {Rix}, {Robin}, {Rocha-Pinto}, {Rockosi}, {Roe},
  {Rollinde}, {Ross}, {Ross}, {Rossetto}, {S{\'a}nchez}, {Santiago}, {Sayres},
  {Schiavon}, {Schlegel}, {Schlesinger}, {Schmidt}, {Schneider}, {Sellgren},
  {Shelden}, {Sheldon}, {Shetrone}, {Shu}, {Silverman}, {Simmerer}, {Simmons},
  {Sivarani}, {Skrutskie}, {Slosar}, {Smee}, {Smith}, {Snedden}, {Stassun},
  {Steele}, {Steinmetz}, {Stockett}, {Stollberg}, {Strauss}, {Szalay},
  {Tanaka}, {Thakar}, {Thomas}, {Tinker}, {Tofflemire}, {Tojeiro}, {Tremonti},
  {Vargas Maga{\~n}a}, {Verde}, {Vogt}, {Wake}, {Wan}, {Wang}, {Weaver},
  {White}, {White}, {Wilson}, {Wisniewski}, {Wood-Vasey}, {Yanny}, {Yasuda},
  {Y{\`e}che}, {York}, {Young}, {Zasowski}, {Zehavi}, \&
  {Zhao}}]{Eisenstein2011}
{Eisenstein}, D.~J., {Weinberg}, D.~H., {Agol}, E., {et~al.} 2011, \aj, 142, 72

\bibitem[{{Fern{\'a}ndez-Trincado}
  {et~al.}(2020{\natexlab{a}}){Fern{\'a}ndez-Trincado}, {Beers}, \&
  {Minniti}}]{Fernandez-Trincado2020a}
{Fern{\'a}ndez-Trincado}, J.~G., {Beers}, T.~C., \& {Minniti}, D.
  2020{\natexlab{a}}, \aap, 644, A83

\bibitem[{{Fern{\'a}ndez-Trincado}
  {et~al.}(2020{\natexlab{b}}){Fern{\'a}ndez-Trincado}, {Beers}, {Minniti},
  {Carigi}, {Barbuy}, {Placco}, {Moni Bidin}, {Villanova}, {Roman-Lopes}, \&
  {Nitschelm}}]{Fernandez-Trincado2020b}
{Fern{\'a}ndez-Trincado}, J.~G., {Beers}, T.~C., {Minniti}, D., {et~al.}
  2020{\natexlab{b}}, \apjl, 903, L17

\bibitem[{{Fern{\'a}ndez-Trincado}
  {et~al.}(2021{\natexlab{a}}){Fern{\'a}ndez-Trincado}, {Beers}, {Minniti},
  {Carigi}, {Placco}, {Chun}, {Lane}, {Geisler}, {Villanova}, {Souza},
  {Barbuy}, {P{\'e}rez-Villegas}, {Chiappini}, {Queiroz}, {Tang},
  {Alonso-Garc{\'\i}a}, {Piatti}, {Palma}, {Alves-Brito}, {Moni Bidin},
  {Roman-Lopes}, {Mu{\~n}oz}, {Singh}, {Kundu}, {Chaves-Velasquez},
  {Romero-Colmenares}, {Longa-Pe{\~n}a}, {Soto}, \&
  {Vieira}}]{Fernandez-Trincado2021b}
{Fern{\'a}ndez-Trincado}, J.~G., {Beers}, T.~C., {Minniti}, D., {et~al.}
  2021{\natexlab{a}}, arXiv e-prints, arXiv:2102.01706

\bibitem[{{Fern{\'a}ndez-Trincado}
  {et~al.}(2020{\natexlab{c}}){Fern{\'a}ndez-Trincado}, {Beers}, {Minniti},
  {Tang}, {Villanova}, {Geisler}, {P{\'e}rez-Villegas}, \&
  {Vieira}}]{Fernandez-Trincado2020c}
{Fern{\'a}ndez-Trincado}, J.~G., {Beers}, T.~C., {Minniti}, D., {et~al.}
  2020{\natexlab{c}}, \aap, 643, L4

\bibitem[{{Fern{\'a}ndez-Trincado}
  {et~al.}(2019{\natexlab{a}}){Fern{\'a}ndez-Trincado}, {Beers}, {Placco},
  {Moreno}, {Alves-Brito}, {Minniti}, {Tang}, {P{\'e}rez-Villegas},
  {Reyl{\'e}}, {Robin}, \& {Villanova}}]{Fernandez-Trincado2019a}
{Fern{\'a}ndez-Trincado}, J.~G., {Beers}, T.~C., {Placco}, V.~M., {et~al.}
  2019{\natexlab{a}}, \apjl, 886, L8

\bibitem[{{Fern{\'a}ndez-Trincado}
  {et~al.}(2019{\natexlab{b}}){Fern{\'a}ndez-Trincado}, {Beers}, {Tang},
  {Moreno}, {P{\'e}rez-Villegas}, \&
  {Ortigoza-Urdaneta}}]{Fernandez-Trincado2019b}
{Fern{\'a}ndez-Trincado}, J.~G., {Beers}, T.~C., {Tang}, B., {et~al.}
  2019{\natexlab{b}}, \mnras, 488, 2864

\bibitem[{{Fern{\'a}ndez-Trincado}
  {et~al.}(2020{\natexlab{d}}){Fern{\'a}ndez-Trincado}, {Chaves-Velasquez},
  {P{\'e}rez-Villegas}, {Vieira}, {Moreno}, {Ortigoza-Urdaneta}, \&
  {Vega-Neme}}]{Fernandez-Trincado2020e}
{Fern{\'a}ndez-Trincado}, J.~G., {Chaves-Velasquez}, L., {P{\'e}rez-Villegas},
  A., {et~al.} 2020{\natexlab{d}}, \mnras, 495, 4113

\bibitem[{{Fern{\'a}ndez-Trincado}
  {et~al.}(2019{\natexlab{c}}){Fern{\'a}ndez-Trincado}, {Mennickent},
  {Cabezas}, {Zamora}, {Martell}, {Beers}, {Placco}, {Nataf},
  {M{\'e}sz{\'a}ros}, {Minniti}, {Schleicher}, {Tang}, {P{\'e}rez-Villegas},
  {Robin}, \& {Reyl{\'e}}}]{Fernandez-Trincado2019c}
{Fern{\'a}ndez-Trincado}, J.~G., {Mennickent}, R., {Cabezas}, M., {et~al.}
  2019{\natexlab{c}}, \aap, 631, A97

\bibitem[{{Fern{\'a}ndez-Trincado}
  {et~al.}(2020{\natexlab{e}}){Fern{\'a}ndez-Trincado}, {Minniti}, {Beers},
  {Villanova}, {Geisler}, {Souza}, {Smith}, {Placco}, {Vieira},
  {P{\'e}rez-Villegas}, {Barbuy}, {Alves-Brito}, {Bidin}, {Alonso-Garc{\'\i}a},
  {Tang}, \& {Palma}}]{Fernandez-Trincado2020d}
{Fern{\'a}ndez-Trincado}, J.~G., {Minniti}, D., {Beers}, T.~C., {et~al.}
  2020{\natexlab{e}}, \aap, 643, A145

\bibitem[{{Fern{\'a}ndez-Trincado}
  {et~al.}(2021{\natexlab{b}}){Fern{\'a}ndez-Trincado}, {Minniti}, {Souza},
  {Beers}, {Geisler}, {Moni Bidin}, {Villanova}, {Majewski}, {Barbuy},
  {P{\'e}rez-Villegas}, {Henao}, {Romero-Colmenares}, {Roman-Lopes}, \&
  {Lane}}]{Fernandez-Trincado2021a}
{Fern{\'a}ndez-Trincado}, J.~G., {Minniti}, D., {Souza}, S.~O., {et~al.}
  2021{\natexlab{b}}, arXiv e-prints, arXiv:2102.01088

\bibitem[{{Fern{\'a}ndez-Trincado} {et~al.}(2016){Fern{\'a}ndez-Trincado},
  {Robin}, {Moreno}, {Schiavon}, {Garc{\'\i}a P{\'e}rez}, {Vieira}, {Cunha},
  {Zamora}, {Sneden}, {Souto}, {Carrera}, {Johnson}, {Shetrone}, {Zasowski},
  {Garc{\'\i}a-Hern{\'a}ndez}, {Majewski}, {Reyl{\'e}}, {Blanco-Cuaresma},
  {Martinez-Medina}, {P{\'e}rez-Villegas}, {Valenzuela}, {Pichardo}, {Meza},
  {M{\'e}sz{\'a}ros}, {Sobeck}, {Geisler}, {Anders}, {Schultheis}, {Tang},
  {Roman-Lopes}, {Mennickent}, {Pan}, {Nitschelm}, \&
  {Allard}}]{Fernandez-Trincado2016b}
{Fern{\'a}ndez-Trincado}, J.~G., {Robin}, A.~C., {Moreno}, E., {et~al.} 2016,
  \apj, 833, 132

\bibitem[{{Fern{\'a}ndez-Trincado} {et~al.}(2015){Fern{\'a}ndez-Trincado},
  {Vivas}, {Mateu}, {Zinn}, {Robin}, {Valenzuela}, {Moreno}, \&
  {Pichardo}}]{Fernandez-Trincado2015a}
{Fern{\'a}ndez-Trincado}, J.~G., {Vivas}, A.~K., {Mateu}, C.~E., {et~al.} 2015,
  \aap, 574, A15

\bibitem[{{Fern{\'a}ndez-Trincado} {et~al.}(2017){Fern{\'a}ndez-Trincado},
  {Zamora}, {Garc{\'\i}a-Hern{\'a}ndez}, {Souto}, {Dell'Agli}, {Schiavon},
  {Geisler}, {Tang}, {Villanova}, {Hasselquist}, {Mennickent}, {Cunha},
  {Shetrone}, {Allende Prieto}, {Vieira}, {Zasowski}, {Sobeck}, {Hayes},
  {Majewski}, {Placco}, {Beers}, {Schleicher}, {Robin}, {M{\'e}sz{\'a}ros},
  {Masseron}, {Garc{\'\i}a P{\'e}rez}, {Anders}, {Meza}, {Alves-Brito},
  {Carrera}, {Minniti}, {Lane}, {Fern{\'a}ndez-Alvar}, {Moreno}, {Pichardo},
  {P{\'e}rez-Villegas}, {Schultheis}, {Roman-Lopes}, {Fuentes}, {Nitschelm},
  {Harding}, {Bizyaev}, {Pan}, {Oravetz}, {Simmons}, {Ivans},
  {Blanco-Cuaresma}, {Hern{\'a}ndez}, {Alonso-Garc{\'\i}a}, {Valenzuela}, \&
  {Chanam{\'e}}}]{Fernandez-Trincado2017}
{Fern{\'a}ndez-Trincado}, J.~G., {Zamora}, O., {Garc{\'\i}a-Hern{\'a}ndez},
  D.~A., {et~al.} 2017, \apjl, 846, L2

\bibitem[{{Fern{\'a}ndez-Trincado}
  {et~al.}(2019{\natexlab{d}}){Fern{\'a}ndez-Trincado}, {Zamora}, {Souto},
  {Cohen}, {Dell'Agli}, {Garc{\'\i}a-Hern{\'a}ndez}, {Masseron}, {Schiavon},
  {M{\'e}sz{\'a}ros}, {Cunha}, {Hasselquist}, {Shetrone}, {Schiappacasse
  Ulloa}, {Tang}, {Geisler}, {Schleicher}, {Villanova}, {Mennickent},
  {Minniti}, {Alonso-Garc{\'\i}a}, {Manchado}, {Beers}, {Sobeck}, {Zasowski},
  {Schultheis}, {Majewski}, {Rojas-Arriagada}, {Almeida}, {Santana}, {Oelkers},
  {Longa-Pe{\~n}a}, {Carrera}, {Burgasser}, {Lane}, {Roman-Lopes}, {Ivans}, \&
  {Hearty}}]{Fernandez-Trincado2019d}
{Fern{\'a}ndez-Trincado}, J.~G., {Zamora}, O., {Souto}, D., {et~al.}
  2019{\natexlab{d}}, \aap, 627, A178

\bibitem[{{Freeman} \& {Bland-Hawthorn}(2002)}]{Freeman2002}
{Freeman}, K. \& {Bland-Hawthorn}, J. 2002, \araa, 40, 487

\bibitem[{{Gaia Collaboration} {et~al.}(2018{\natexlab{a}}){Gaia
  Collaboration}, {Brown}, {Vallenari}, {Prusti}, {de Bruijne}, {Babusiaux},
  {Bailer-Jones}, {Biermann}, {Evans}, {Eyer}, {Jansen}, {Jordi}, {Klioner},
  {Lammers}, {Lindegren}, {Luri}, {Mignard}, {Panem}, {Pourbaix}, {Randich},
  {Sartoretti}, {Siddiqui}, {Soubiran}, {van Leeuwen}, {Walton}, {Arenou},
  {Bastian}, {Cropper}, {Drimmel}, {Katz}, {Lattanzi}, {Bakker}, {Cacciari},
  {Casta{\~n}eda}, {Chaoul}, {Cheek}, {De Angeli}, {Fabricius}, {Guerra},
  {Holl}, {Masana}, {Messineo}, {Mowlavi}, {Nienartowicz}, {Panuzzo},
  {Portell}, {Riello}, {Seabroke}, {Tanga}, {Th{\'e}venin}, {Gracia-Abril},
  {Comoretto}, {Garcia-Reinaldos}, {Teyssier}, {Altmann}, {Andrae}, {Audard},
  {Bellas-Velidis}, {Benson}, {Berthier}, {Blomme}, {Burgess}, {Busso},
  {Carry}, {Cellino}, {Clementini}, {Clotet}, {Creevey}, {Davidson}, {De
  Ridder}, {Delchambre}, {Dell'Oro}, {Ducourant},
  {Fern{\'a}ndez-Hern{\'a}ndez}, {Fouesneau}, {Fr{\'e}mat}, {Galluccio},
  {Garc{\'\i}a-Torres}, {Gonz{\'a}lez-N{\'u}{\~n}ez}, {Gonz{\'a}lez-Vidal},
  {Gosset}, {Guy}, {Halbwachs}, {Hambly}, {Harrison}, {Hern{\'a}ndez},
  {Hestroffer}, {Hodgkin}, {Hutton}, {Jasniewicz}, {Jean-Antoine-Piccolo},
  {Jordan}, {Korn}, {Krone-Martins}, {Lanzafame}, {Lebzelter}, {L{\"o}ffler},
  {Manteiga}, {Marrese}, {Mart{\'\i}n-Fleitas}, {Moitinho}, {Mora}, {Muinonen},
  {Osinde}, {Pancino}, {Pauwels}, {Petit}, {Recio-Blanco}, {Richards},
  {Rimoldini}, {Robin}, {Sarro}, {Siopis}, {Smith}, {Sozzetti}, {S{\"u}veges},
  {Torra}, {van Reeven}, {Abbas}, {Abreu Aramburu}, {Accart}, {Aerts},
  {Altavilla}, {{\'A}lvarez}, {Alvarez}, {Alves}, {Anderson}, {Andrei},
  {Anglada Varela}, {Antiche}, {Antoja}, {Arcay}, {Astraatmadja}, {Bach},
  {Baker}, {Balaguer-N{\'u}{\~n}ez}, {Balm}, {Barache}, {Barata}, {Barbato},
  {Barblan}, {Barklem}, {Barrado}, {Barros}, {Barstow}, {Bartholom{\'e}
  Mu{\~n}oz}, {Bassilana}, {Becciani}, {Bellazzini}, {Berihuete}, {Bertone},
  {Bianchi}, {Bienaym{\'e}}, {Blanco-Cuaresma}, {Boch}, {Boeche}, {Bombrun},
  {Borrachero}, {Bossini}, {Bouquillon}, {Bourda}, {Bragaglia}, {Bramante},
  {Breddels}, {Bressan}, {Brouillet}, {Br{\"u}semeister}, {Brugaletta},
  {Bucciarelli}, {Burlacu}, {Busonero}, {Butkevich}, {Buzzi}, {Caffau},
  {Cancelliere}, {Cannizzaro}, {Cantat-Gaudin}, {Carballo}, {Carlucci},
  {Carrasco}, {Casamiquela}, {Castellani}, {Castro-Ginard}, {Charlot},
  {Chemin}, {Chiavassa}, {Cocozza}, {Costigan}, {Cowell}, {Crifo}, {Crosta},
  {Crowley}, {Cuypers}, {Dafonte}, {Damerdji}, {Dapergolas}, {David}, {David},
  {de Laverny}, {De Luise}, {De March}, {de Martino}, {de Souza}, {de Torres},
  {Debosscher}, {del Pozo}, {Delbo}, {Delgado}, {Delgado}, {Di Matteo},
  {Diakite}, {Diener}, {Distefano}, {Dolding}, {Drazinos}, {Dur{\'a}n},
  {Edvardsson}, {Enke}, {Eriksson}, {Esquej}, {Eynard Bontemps}, {Fabre},
  {Fabrizio}, {Faigler}, {Falc{\~a}o}, {Farr{\`a}s Casas}, {Federici},
  {Fedorets}, {Fernique}, {Figueras}, {Filippi}, {Findeisen}, {Fonti},
  {Fraile}, {Fraser}, {Fr{\'e}zouls}, {Gai}, {Galleti}, {Garabato},
  {Garc{\'\i}a-Sedano}, {Garofalo}, {Garralda}, {Gavel}, {Gavras}, {Gerssen},
  {Geyer}, {Giacobbe}, {Gilmore}, {Girona}, {Giuffrida}, {Glass}, {Gomes},
  {Granvik}, {Gueguen}, {Guerrier}, {Guiraud}, {Guti{\'e}rrez-S{\'a}nchez},
  {Haigron}, {Hatzidimitriou}, {Hauser}, {Haywood}, {Heiter}, {Helmi}, {Heu},
  {Hilger}, {Hobbs}, {Hofmann}, {Holland}, {Huckle}, {Hypki}, {Icardi},
  {Jan{\ss}en}, {Jevardat de Fombelle}, {Jonker}, {Juh{\'a}sz}, {Julbe},
  {Karampelas}, {Kewley}, {Klar}, {Kochoska}, {Kohley}, {Kolenberg},
  {Kontizas}, {Kontizas}, {Koposov}, {Kordopatis}, {Kostrzewa-Rutkowska},
  {Koubsky}, {Lambert}, {Lanza}, {Lasne}, {Lavigne}, {Le Fustec}, {Le
  Poncin-Lafitte}, {Lebreton}, {Leccia}, {Leclerc}, {Lecoeur-Taibi},
  {Lenhardt}, {Leroux}, {Liao}, {Licata}, {Lindstr{\o}m}, {Lister}, {Livanou},
  {Lobel}, {L{\'o}pez}, {Managau}, {Mann}, {Mantelet}, {Marchal}, {Marchant},
  {Marconi}, {Marinoni}, {Marschalk{\'o}}, {Marshall}, {Martino}, {Marton},
  {Mary}, {Massari}, {Matijevi{\v{c}}}, {Mazeh}, {McMillan}, {Messina},
  {Michalik}, {Millar}, {Molina}, {Molinaro}, {Moln{\'a}r}, {Montegriffo},
  {Mor}, {Morbidelli}, {Morel}, {Morris}, {Mulone}, {Muraveva}, {Musella},
  {Nelemans}, {Nicastro}, {Noval}, {O'Mullane}, {Ord{\'e}novic},
  {Ord{\'o}{\~n}ez-Blanco}, {Osborne}, {Pagani}, {Pagano}, {Pailler},
  {Palacin}, {Palaversa}, {Panahi}, {Pawlak}, {Piersimoni}, {Pineau}, {Plachy},
  {Plum}, {Poggio}, {Poujoulet}, {Pr{\v{s}}a}, {Pulone}, {Racero}, {Ragaini},
  {Rambaux}, {Ramos-Lerate}, {Regibo}, {Reyl{\'e}}, {Riclet}, {Ripepi}, {Riva},
  {Rivard}, {Rixon}, {Roegiers}, {Roelens}, {Romero-G{\'o}mez}, {Rowell},
  {Royer}, {Ruiz-Dern}, {Sadowski}, {Sagrist{\`a} Sell{\'e}s}, {Sahlmann},
  {Salgado}, {Salguero}, {Sanna}, {Santana-Ros}, {Sarasso}, {Savietto},
  {Schultheis}, {Sciacca}, {Segol}, {Segovia}, {S{\'e}gransan}, {Shih},
  {Siltala}, {Silva}, {Smart}, {Smith}, {Solano}, {Solitro}, {Sordo}, {Soria
  Nieto}, {Souchay}, {Spagna}, {Spoto}, {Stampa}, {Steele},
  {Steidelm{\"u}ller}, {Stephenson}, {Stoev}, {Suess}, {Surdej}, {Szabados},
  {Szegedi-Elek}, {Tapiador}, {Taris}, {Tauran}, {Taylor}, {Teixeira},
  {Terrett}, {Teyssand ier}, {Thuillot}, {Titarenko}, {Torra Clotet}, {Turon},
  {Ulla}, {Utrilla}, {Uzzi}, {Vaillant}, {Valentini}, {Valette}, {van Elteren},
  {Van Hemelryck}, {van Leeuwen}, {Vaschetto}, {Vecchiato}, {Veljanoski},
  {Viala}, {Vicente}, {Vogt}, {von Essen}, {Voss}, {Votruba}, {Voutsinas},
  {Walmsley}, {Weiler}, {Wertz}, {Wevers}, {Wyrzykowski}, {Yoldas},
  {{\v{Z}}erjal}, {Ziaeepour}, {Zorec}, {Zschocke}, {Zucker}, {Zurbach}, \&
  {Zwitter}}]{Brown2018}
{Gaia Collaboration}, {Brown}, A.~G.~A., {Vallenari}, A., {et~al.}
  2018{\natexlab{a}}, \aap, 616, A1

\bibitem[{{Gaia Collaboration} {et~al.}(2020){Gaia Collaboration}, {Brown},
  {Vallenari}, {Prusti}, {de Bruijne}, {Babusiaux}, \& {Biermann}}]{Brown2020}
{Gaia Collaboration}, {Brown}, A.~G.~A., {Vallenari}, A., {et~al.} 2020, arXiv
  e-prints, arXiv:2012.01533

\bibitem[{{Gaia Collaboration} {et~al.}(2018{\natexlab{b}}){Gaia
  Collaboration}, {Helmi}, {van Leeuwen}, {McMillan}, {Massari}, {Antoja},
  {Robin}, {Lindegren}, {Bastian}, {Arenou}, {Babusiaux}, {Biermann},
  {Breddels}, {Hobbs}, {Jordi}, {Pancino}, {Reyl{\'e}}, {Veljanoski}, {Brown},
  {Vallenari}, {Prusti}, {de Bruijne}, {Bailer-Jones}, {Evans}, {Eyer},
  {Jansen}, {Klioner}, {Lammers}, {Luri}, {Mignard}, {Panem}, {Pourbaix},
  {Randich}, {Sartoretti}, {Siddiqui}, {Soubiran}, {Walton}, {Cropper},
  {Drimmel}, {Katz}, {Lattanzi}, {Bakker}, {Cacciari}, {Casta{\~n}eda},
  {Chaoul}, {Cheek}, {De Angeli}, {Fabricius}, {Guerra}, {Holl}, {Masana},
  {Messineo}, {Mowlavi}, {Nienartowicz}, {Panuzzo}, {Portell}, {Riello},
  {Seabroke}, {Tanga}, {Th{\'e}venin}, {Gracia-Abril}, {Comoretto},
  {Garcia-Reinaldos}, {Teyssier}, {Altmann}, {Andrae}, {Audard},
  {Bellas-Velidis}, {Benson}, {Berthier}, {Blomme}, {Burgess}, {Busso},
  {Carry}, {Cellino}, {Clementini}, {Clotet}, {Creevey}, {Davidson}, {De
  Ridder}, {Delchambre}, {Dell'Oro}, {Ducourant},
  {Fern{\'a}ndez-Hern{\'a}ndez}, {Fouesneau}, {Fr{\'e}mat}, {Galluccio},
  {Garc{\'\i}a-Torres}, {Gonz{\'a}lez-N{\'u}{\~n}ez}, {Gonz{\'a}lez-Vidal},
  {Gosset}, {Guy}, {Halbwachs}, {Hambly}, {Harrison}, {Hern{\'a}ndez},
  {Hestroffer}, {Hodgkin}, {Hutton}, {Jasniewicz}, {Jean-Antoine-Piccolo},
  {Jordan}, {Korn}, {Krone-Martins}, {Lanzafame}, {Lebzelter}, {L{\"o}ffler},
  {Manteiga}, {Marrese}, {Mart{\'\i}n-Fleitas}, {Moitinho}, {Mora}, {Muinonen},
  {Osinde}, {Pauwels}, {Petit}, {Recio-Blanco}, {Richards}, {Rimoldini},
  {Sarro}, {Siopis}, {Smith}, {Sozzetti}, {S{\"u}veges}, {Torra}, {van Reeven},
  {Abbas}, {Abreu Aramburu}, {Accart}, {Aerts}, {Altavilla}, {{\'A}lvarez},
  {Alvarez}, {Alves}, {Anderson}, {Andrei}, {Anglada Varela}, {Antiche},
  {Arcay}, {Astraatmadja}, {Bach}, {Baker}, {Balaguer-N{\'u}{\~n}ez}, {Balm},
  {Barache}, {Barata}, {Barbato}, {Barblan}, {Barklem}, {Barrado}, {Barros},
  {Barstow}, {Bartholom{\'e} Mu{\~n}oz}, {Bassilana}, {Becciani}, {Bellazzini},
  {Berihuete}, {Bertone}, {Bianchi}, {Bienaym{\'e}}, {Blanco-Cuaresma}, {Boch},
  {Boeche}, {Bombrun}, {Borrachero}, {Bossini}, {Bouquillon}, {Bourda},
  {Bragaglia}, {Bramante}, {Bressan}, {Brouillet}, {Br{\"u}semeister},
  {Brugaletta}, {Bucciarelli}, {Burlacu}, {Busonero}, {Butkevich}, {Buzzi},
  {Caffau}, {Cancelliere}, {Cannizzaro}, {Cantat-Gaudin}, {Carballo},
  {Carlucci}, {Carrasco}, {Casamiquela}, {Castellani}, {Castro-Ginard},
  {Charlot}, {Chemin}, {Chiavassa}, {Cocozza}, {Costigan}, {Cowell}, {Crifo},
  {Crosta}, {Crowley}, {Cuypers}, {Dafonte}, {Damerdji}, {Dapergolas}, {David},
  {David}, {de Laverny}, {De Luise}, {De March}, {de Martino}, {de Souza}, {de
  Torres}, {Debosscher}, {del Pozo}, {Delbo}, {Delgado}, {Delgado}, {Di
  Matteo}, {Diakite}, {Diener}, {Distefano}, {Dolding}, {Drazinos},
  {Dur{\'a}n}, {Edvardsson}, {Enke}, {Eriksson}, {Esquej}, {Eynard Bontemps},
  {Fabre}, {Fabrizio}, {Faigler}, {Falc{\~a}o}, {Farr{\`a}s Casas}, {Federici},
  {Fedorets}, {Fernique}, {Figueras}, {Filippi}, {Findeisen}, {Fonti},
  {Fraile}, {Fraser}, {Fr{\'e}zouls}, {Gai}, {Galleti}, {Garabato},
  {Garc{\'\i}a-Sedano}, {Garofalo}, {Garralda}, {Gavel}, {Gavras}, {Gerssen},
  {Geyer}, {Giacobbe}, {Gilmore}, {Girona}, {Giuffrida}, {Glass}, {Gomes},
  {Granvik}, {Gueguen}, {Guerrier}, {Guiraud}, {Guti{\'e}rrez-S{\'a}nchez},
  {Hofmann}, {Holland}, {Huckle}, {Hypki}, {Icardi}, {Jan{\ss}en}, {Jevardat de
  Fombelle}, {Jonker}, {Juh{\'a}sz}, {Julbe}, {Karampelas}, {Kewley}, {Klar},
  {Kochoska}, {Kohley}, {Kolenberg}, {Kontizas}, {Kontizas}, {Koposov},
  {Kordopatis}, {Kostrzewa-Rutkowska}, {Koubsky}, {Lambert}, {Lanza}, {Lasne},
  {Lavigne}, {Le Fustec}, {Le Poncin-Lafitte}, {Lebreton}, {Leccia}, {Leclerc},
  {Lecoeur-Taibi}, {Lenhardt}, {Leroux}, {Liao}, {Licata}, {Lindstr{\o}m},
  {Lister}, {Livanou}, {Lobel}, {L{\'o}pez}, {Managau}, {Mann}, {Mantelet},
  {Marchal}, {Marchant}, {Marconi}, {Marinoni}, {Marschalk{\'o}}, {Marshall},
  {Martino}, {Marton}, {Mary}, {Matijevi{\v{c}}}, {Mazeh}, {Messina},
  {Michalik}, {Millar}, {Molina}, {Molinaro}, {Moln{\'a}r}, {Montegriffo},
  {Mor}, {Morbidelli}, {Morel}, {Morris}, {Mulone}, {Muraveva}, {Musella},
  {Nelemans}, {Nicastro}, {Noval}, {O'Mullane}, {Ord{\'e}novic},
  {Ord{\'o}{\~n}ez-Blanco}, {Osborne}, {Pagani}, {Pagano}, {Pailler},
  {Palacin}, {Palaversa}, {Panahi}, {Pawlak}, {Piersimoni}, {Pineau}, {Plachy},
  {Plum}, {Poggio}, {Poujoulet}, {Pr{\v{s}}a}, {Pulone}, {Racero}, {Ragaini},
  {Rambaux}, {Ramos-Lerate}, {Regibo}, {Riclet}, {Ripepi}, {Riva}, {Rivard},
  {Rixon}, {Roegiers}, {Roelens}, {Romero-G{\'o}mez}, {Rowell}, {Royer},
  {Ruiz-Dern}, {Sadowski}, {Sagrist{\`a} Sell{\'e}s}, {Sahlmann}, {Salgado},
  {Salguero}, {Sanna}, {Santana-Ros}, {Sarasso}, {Savietto}, {Schultheis},
  {Sciacca}, {Segol}, {Segovia}, {S{\'e}gransan}, {Shih}, {Siltala}, {Silva},
  {Smart}, {Smith}, {Solano}, {Solitro}, {Sordo}, {Soria Nieto}, {Souchay},
  {Spagna}, {Spoto}, {Stampa}, {Steele}, {Steidelm{\"u}ller}, {Stephenson},
  {Stoev}, {Suess}, {Surdej}, {Szabados}, {Szegedi-Elek}, {Tapiador}, {Taris},
  {Tauran}, {Taylor}, {Teixeira}, {Terrett}, {Teyssandier}, {Thuillot},
  {Titarenko}, {Torra Clotet}, {Turon}, {Ulla}, {Utrilla}, {Uzzi}, {Vaillant},
  {Valentini}, {Valette}, {van Elteren}, {Van Hemelryck}, {van Leeuwen},
  {Vaschetto}, {Vecchiato}, {Viala}, {Vicente}, {Vogt}, {von Essen}, {Voss},
  {Votruba}, {Voutsinas}, {Walmsley}, {Weiler}, {Wertz}, {Wevems},
  {Wyrzykowski}, {Yoldas}, {{\v{Z}}erjal}, {Ziaeepour}, {Zorec}, {Zschocke},
  {Zucker}, {Zurbach}, \& {Zwitter}}]{Helmi2018}
{Gaia Collaboration}, {Helmi}, A., {van Leeuwen}, F., {et~al.}
  2018{\natexlab{b}}, \aap, 616, A12

\bibitem[{{Garc{\'\i}a P{\'e}rez} {et~al.}(2016){Garc{\'\i}a P{\'e}rez},
  {Allende Prieto}, {Holtzman}, {Shetrone}, {M{\'e}sz{\'a}ros}, {Bizyaev},
  {Carrera}, {Cunha}, {Garc{\'\i}a-Hern{\'a}ndez}, {Johnson}, {Majewski},
  {Nidever}, {Schiavon}, {Shane}, {Smith}, {Sobeck}, {Troup}, {Zamora},
  {Weinberg}, {Bovy}, {Eisenstein}, {Feuillet}, {Frinchaboy}, {Hayden},
  {Hearty}, {Nguyen}, {O'Connell}, {Pinsonneault}, {Wilson}, \&
  {Zasowski}}]{Garcia2016}
{Garc{\'\i}a P{\'e}rez}, A.~E., {Allende Prieto}, C., {Holtzman}, J.~A.,
  {et~al.} 2016, \aj, 151, 144

\bibitem[{{Gunn} {et~al.}(2006){Gunn}, {Siegmund}, {Mannery}, {Owen}, {Hull},
  {Leger}, {Carey}, {Knapp}, {York}, {Boroski}, {Kent}, {Lupton}, {Rockosi},
  {Evans}, {Waddell}, {Anderson}, {Annis}, {Barentine}, {Bartoszek}, {Bastian},
  {Bracker}, {Brewington}, {Briegel}, {Brinkmann}, {Brown}, {Carr},
  {Czarapata}, {Drennan}, {Dombeck}, {Federwitz}, {Gillespie}, {Gonzales},
  {Hansen}, {Harvanek}, {Hayes}, {Jordan}, {Kinney}, {Klaene}, {Kleinman},
  {Kron}, {Kresinski}, {Lee}, {Limmongkol}, {Lindenmeyer}, {Long}, {Loomis},
  {McGehee}, {Mantsch}, {Neilsen}, {Neswold}, {Newman}, {Nitta}, {Peoples},
  {Pier}, {Prieto}, {Prosapio}, {Rivetta}, {Schneider}, {Snedden}, \&
  {Wang}}]{Gunn2006}
{Gunn}, J.~E., {Siegmund}, W.~A., {Mannery}, E.~J., {et~al.} 2006, \aj, 131,
  2332

\bibitem[{{Gustafsson} {et~al.}(2008){Gustafsson}, {Edvardsson}, {Eriksson},
  {J{\o}rgensen}, {Nordlund}, \& {Plez}}]{Gustafsson2008}
{Gustafsson}, B., {Edvardsson}, B., {Eriksson}, K., {et~al.} 2008, \aap, 486,
  951

\bibitem[{{Hanke} {et~al.}(2020){Hanke}, {Koch}, {Prudil}, {Grebel}, \&
  {Bastian}}]{Hanke2020}
{Hanke}, M., {Koch}, A., {Prudil}, Z., {Grebel}, E.~K., \& {Bastian}, U. 2020,
  \aap, 637, A98

\bibitem[{{Harris}(1996)}]{Harris1996}
{Harris}, W.~E. 1996, \aj, 112, 1487

\bibitem[{{Hasselquist} {et~al.}(2019){Hasselquist}, {Carlin}, {Holtzman},
  {Shetrone}, {Hayes}, {Cunha}, {Smith}, {Beaton}, {Sobeck}, {Allende Prieto},
  {Majewski}, {Anguiano}, {Bizyaev}, {Garc{\'\i}a-Hern{\'a}ndez}, {Lane},
  {Pan}, {Nidever}, {Fern{\'a}ndez-Trincado}, {Wilson}, \&
  {Zamora}}]{Hasselquist2019}
{Hasselquist}, S., {Carlin}, J.~L., {Holtzman}, J.~A., {et~al.} 2019, \apj,
  872, 58

\bibitem[{{Hasselquist} {et~al.}(2016){Hasselquist}, {Shetrone}, {Cunha},
  {Smith}, {Holtzman}, {Lawler}, {Allende Prieto}, {Beers}, {Chojnowski},
  {Fern{\'a}ndez-Trincado}, {Garc{\'\i}a-Hern{\'a}ndez}, {Hearty}, {Majewski},
  {Pereira}, {Placco}, {Villanova}, \& {Zamora}}]{Hasselquist2016}
{Hasselquist}, S., {Shetrone}, M., {Cunha}, K., {et~al.} 2016, \apj, 833, 81

\bibitem[{{Hasselquist} {et~al.}(2017){Hasselquist}, {Shetrone}, {Smith},
  {Holtzman}, {McWilliam}, {Fern{\'a}ndez-Trincado}, {Beers}, {Majewski},
  {Nidever}, {Tang}, {Tissera}, {Fern{\'a}ndez Alvar}, {Allende Prieto},
  {Almeida}, {Anguiano}, {Battaglia}, {Carigi}, {Delgado Inglada},
  {Frinchaboy}, {Garc{\'\i}a-Hern{\'a}ndez}, {Geisler}, {Minniti}, {Placco},
  {Schultheis}, {Sobeck}, \& {Villanova}}]{Hasselquist2017}
{Hasselquist}, S., {Shetrone}, M., {Smith}, V., {et~al.} 2017, \apj, 845, 162

\bibitem[{{Hayes} {et~al.}(2020){Hayes}, {Majewski}, {Hasselquist}, {Anguiano},
  {Shetrone}, {Law}, {Schiavon}, {Cunha}, {Smith}, {Beaton}, {Price-Whelan},
  {Allende Prieto}, {Battaglia}, {Bizyaev}, {Brownstein}, {Cohen},
  {Frinchaboy}, {Garc{\'\i}a-Hern{\'a}ndez}, {Lacerna}, {Lane},
  {M{\'e}sz{\'a}ros}, {Bidin}, {M{\~{u}}noz}, {Nidever}, {Oravetz}, {Oravetz},
  {Pan}, {Roman-Lopes}, {Sobeck}, \& {Stringfellow}}]{Hayes2020}
{Hayes}, C.~R., {Majewski}, S.~R., {Hasselquist}, S., {et~al.} 2020, \apj, 889,
  63

\bibitem[{{Holtzman} {et~al.}(2018){Holtzman}, {Hasselquist}, {Shetrone},
  {Cunha}, {Allende Prieto}, {Anguiano}, {Bizyaev}, {Bovy}, {Casey},
  {Edvardsson}, {Johnson}, {J{\"o}nsson}, {Meszaros}, {Smith}, {Sobeck},
  {Zamora}, {Chojnowski}, {Fernandez-Trincado}, {Garcia-Hernandez}, {Majewski},
  {Pinsonneault}, {Souto}, {Stringfellow}, {Tayar}, {Troup}, \&
  {Zasowski}}]{Holtzman2018}
{Holtzman}, J.~A., {Hasselquist}, S., {Shetrone}, M., {et~al.} 2018, \aj, 156,
  125

\bibitem[{{Holtzman} {et~al.}(2015){Holtzman}, {Shetrone}, {Johnson}, {Allende
  Prieto}, {Anders}, {Andrews}, {Beers}, {Bizyaev}, {Blanton}, {Bovy},
  {Carrera}, {Chojnowski}, {Cunha}, {Eisenstein}, {Feuillet}, {Frinchaboy},
  {Galbraith-Frew}, {Garc{\'\i}a P{\'e}rez}, {Garc{\'\i}a-Hern{\'a}ndez},
  {Hasselquist}, {Hayden}, {Hearty}, {Ivans}, {Majewski}, {Martell},
  {Meszaros}, {Muna}, {Nidever}, {Nguyen}, {O'Connell}, {Pan}, {Pinsonneault},
  {Robin}, {Schiavon}, {Shane}, {Sobeck}, {Smith}, {Troup}, {Weinberg},
  {Wilson}, {Wood-Vasey}, {Zamora}, \& {Zasowski}}]{Holtzman2015}
{Holtzman}, J.~A., {Shetrone}, M., {Johnson}, J.~A., {et~al.} 2015, \aj, 150,
  148

\bibitem[{{Horta} {et~al.}(2021){Horta}, {Mackereth}, {Schiavon},
  {Hasselquist}, {Bovy}, {Allende Prieto}, {Beers}, {Cunha},
  {Garc{\'\i}a-Hern{\'a}ndez}, {Kisku}, {Lane}, {Majewski}, {Mason}, {Nataf},
  {Roman-Lopes}, \& {Schultheis}}]{Horta2021}
{Horta}, D., {Mackereth}, J.~T., {Schiavon}, R.~P., {et~al.} 2021, \mnras, 500,
  5462

\bibitem[{{Huang} \& {Koposov}(2020)}]{Huang2020}
{Huang}, K.-W. \& {Koposov}, S.~E. 2020, \mnras [\eprint[arXiv]{2005.14014}]

\bibitem[{{Ibata} {et~al.}(2001){Ibata}, {Irwin}, {Lewis}, \&
  {Stolte}}]{Ibata2001}
{Ibata}, R., {Irwin}, M., {Lewis}, G.~F., \& {Stolte}, A. 2001, \apjl, 547,
  L133

\bibitem[{{Ibata} {et~al.}(1994){Ibata}, {Gilmore}, \& {Irwin}}]{Ibata1994}
{Ibata}, R.~A., {Gilmore}, G., \& {Irwin}, M.~J. 1994, \nat, 370, 194

\bibitem[{{J{\"o}nsson} {et~al.}(2018){J{\"o}nsson}, {Allende Prieto},
  {Holtzman}, {Feuillet}, {Hawkins}, {Cunha}, {M{\'e}sz{\'a}ros},
  {Hasselquist}, {Fern{\'a}ndez-Trincado}, {Garc{\'\i}a-Hern{\'a}ndez},
  {Bizyaev}, {Carrera}, {Majewski}, {Pinsonneault}, {Shetrone}, {Smith},
  {Sobeck}, {Souto}, {Stringfellow}, {Teske}, \& {Zamora}}]{Henrik2018}
{J{\"o}nsson}, H., {Allende Prieto}, C., {Holtzman}, J.~A., {et~al.} 2018, \aj,
  156, 126

\bibitem[{{J{\"o}nsson} {et~al.}(2020){J{\"o}nsson}, {Holtzman}, {Allende
  Prieto}, {Cunha}, {Garc{\'\i}a-Hern{\'a}ndez}, {Hasselquist}, {Masseron},
  {Osorio}, {Shetrone}, {Smith}, {Stringfellow}, {Bizyaev}, {Edvardsson},
  {Majewski}, {M{\'e}sz{\'a}ros}, {Souto}, {Zamora}, {Beaton}, {Bovy}, {Donor},
  {Pinsonneault}, {Poovelil}, \& {Sobeck}}]{Henrik2020}
{J{\"o}nsson}, H., {Holtzman}, J.~A., {Allende Prieto}, C., {et~al.} 2020, \aj,
  160, 120

\bibitem[{{Karlsson} {et~al.}(2012){Karlsson}, {Bland-Hawthorn}, {Freeman}, \&
  {Silk}}]{Karlsson2012}
{Karlsson}, T., {Bland-Hawthorn}, J., {Freeman}, K.~C., \& {Silk}, J. 2012,
  \apj, 759, 111

\bibitem[{{Law} {et~al.}(2005){Law}, {Johnston}, \& {Majewski}}]{Law2005}
{Law}, D.~R., {Johnston}, K.~V., \& {Majewski}, S.~R. 2005, \apj, 619, 807

\bibitem[{{Law} \& {Majewski}(2010)}]{Law2010a}
{Law}, D.~R. \& {Majewski}, S.~R. 2010, \apj, 718, 1128

\bibitem[{{Li} {et~al.}(2019){Li}, {FELLOW}, {Liu}, {Xue}, {Zhong}, {Weiss},
  {Carlin}, {Tian}, \& {FELLOW}}]{Li2019}
{Li}, J., {FELLOW}, L., {Liu}, C., {et~al.} 2019, \apj, 874, 138

\bibitem[{{Majewski} {et~al.}(2017){Majewski}, {Schiavon}, {Frinchaboy},
  {Allende Prieto}, {Barkhouser}, {Bizyaev}, {Blank}, {Brunner}, {Burton},
  {Carrera}, {Chojnowski}, {Cunha}, {Epstein}, {Fitzgerald}, {Garc{\'\i}a
  P{\'e}rez}, {Hearty}, {Henderson}, {Holtzman}, {Johnson}, {Lam}, {Lawler},
  {Maseman}, {M{\'e}sz{\'a}ros}, {Nelson}, {Nguyen}, {Nidever}, {Pinsonneault},
  {Shetrone}, {Smee}, {Smith}, {Stolberg}, {Skrutskie}, {Walker}, {Wilson},
  {Zasowski}, {Anders}, {Basu}, {Beland}, {Blanton}, {Bovy}, {Brownstein},
  {Carlberg}, {Chaplin}, {Chiappini}, {Eisenstein}, {Elsworth}, {Feuillet},
  {Fleming}, {Galbraith-Frew}, {Garc{\'\i}a}, {Garc{\'\i}a-Hern{\'a}ndez},
  {Gillespie}, {Girardi}, {Gunn}, {Hasselquist}, {Hayden}, {Hekker}, {Ivans},
  {Kinemuchi}, {Klaene}, {Mahadevan}, {Mathur}, {Mosser}, {Muna}, {Munn},
  {Nichol}, {O'Connell}, {Parejko}, {Robin}, {Rocha-Pinto}, {Schultheis},
  {Serenelli}, {Shane}, {Silva Aguirre}, {Sobeck}, {Thompson}, {Troup},
  {Weinberg}, \& {Zamora}}]{Majewski2017}
{Majewski}, S.~R., {Schiavon}, R.~P., {Frinchaboy}, P.~M., {et~al.} 2017, \aj,
  154, 94

\bibitem[{{Majewski} {et~al.}(2003){Majewski}, {Skrutskie}, {Weinberg}, \&
  {Ostheimer}}]{Majewski2003}
{Majewski}, S.~R., {Skrutskie}, M.~F., {Weinberg}, M.~D., \& {Ostheimer}, J.~C.
  2003, \apj, 599, 1082

\bibitem[{{Massari} {et~al.}(2019){Massari}, {Koppelman}, \&
  {Helmi}}]{Massari2019}
{Massari}, D., {Koppelman}, H.~H., \& {Helmi}, A. 2019, \aap, 630, L4

\bibitem[{{Masseron} {et~al.}(2016){Masseron}, {Merle}, \&
  {Hawkins}}]{Masseron2016}
{Masseron}, T., {Merle}, T., \& {Hawkins}, K. 2016, {BACCHUS: Brussels
  Automatic Code for Characterizing High accUracy Spectra}

\bibitem[{{Mateo} {et~al.}(1996){Mateo}, {Mirabal}, {Udalski}, {Szymanski},
  {Kaluzny}, {Kubiak}, {Krzeminski}, \& {Stanek}}]{Mateo1996}
{Mateo}, M., {Mirabal}, N., {Udalski}, A., {et~al.} 1996, \apjl, 458, L13

\bibitem[{{McWilliam} {et~al.}(2013){McWilliam}, {Wallerstein}, \&
  {Mottini}}]{McWilliam2013}
{McWilliam}, A., {Wallerstein}, G., \& {Mottini}, M. 2013, \apj, 778, 149

\bibitem[{{M{\'e}sz{\'a}ros} {et~al.}(2020){M{\'e}sz{\'a}ros}, {Masseron},
  {Garc{\'\i}a-Hern{\'a}ndez}, {Allende Prieto}, {Beers}, {Bizyaev},
  {Chojnowski}, {Cohen}, {Cunha}, {Dell'Agli}, {Ebelke},
  {Fern{\'a}ndez-Trincado}, {Frinchaboy}, {Geisler}, {Hasselquist}, {Hearty},
  {Holtzman}, {Johnson}, {Lane}, {Lacerna}, {Longa-Pe{\~n}a}, {Majewski},
  {Martell}, {Minniti}, {Nataf}, {Nidever}, {Pan}, {Schiavon}, {Shetrone},
  {Smith}, {Sobeck}, {Stringfellow}, {Szigeti}, {Tang}, {Wilson}, \&
  {Zamora}}]{Meszaros2020}
{M{\'e}sz{\'a}ros}, S., {Masseron}, T., {Garc{\'\i}a-Hern{\'a}ndez}, D.~A.,
  {et~al.} 2020, \mnras, 492, 1641

\bibitem[{{Milone} {et~al.}(2017){Milone}, {Piotto}, {Renzini}, {Marino},
  {Bedin}, {Vesperini}, {D'Antona}, {Nardiello}, {Anderson}, {King}, {Yong},
  {Bellini}, {Aparicio}, {Barbuy}, {Brown}, {Cassisi}, {Ortolani}, {Salaris},
  {Sarajedini}, \& {van der Marel}}]{Milone2017}
{Milone}, A.~P., {Piotto}, G., {Renzini}, A., {et~al.} 2017, \mnras, 464, 3636

\bibitem[{{Nataf} {et~al.}(2019){Nataf}, {Wyse}, {Schiavon}, {Ting}, {Minniti},
  {Cohen}, {Fern{\'a}ndez-Trincado}, {Geisler}, {Nitschelm}, \&
  {Frinchaboy}}]{Nataf2019}
{Nataf}, D.~M., {Wyse}, R. F.~G., {Schiavon}, R.~P., {et~al.} 2019, \aj, 158,
  14

\bibitem[{{Newberg} {et~al.}(2003){Newberg}, {Yanny}, {Grebel}, {Hennessy},
  {Ivezi{\'c}}, {Martinez-Delgado}, {Odenkirchen}, {Rix}, {Brinkmann}, {Lamb},
  {Schneider}, \& {York}}]{Newberg2003}
{Newberg}, H.~J., {Yanny}, B., {Grebel}, E.~K., {et~al.} 2003, \apjl, 596, L191

\bibitem[{{Nidever} {et~al.}(2015){Nidever}, {Holtzman}, {Allende Prieto},
  {Beland}, {Bender}, {Bizyaev}, {Burton}, {Desphande}, {Fleming}, {Garc{\'\i}a
  P{\'e}rez}, {Hearty}, {Majewski}, {M{\'e}sz{\'a}ros}, {Muna}, {Nguyen},
  {Schiavon}, {Shetrone}, {Skrutskie}, {Sobeck}, \& {Wilson}}]{Nidever2015}
{Nidever}, D.~L., {Holtzman}, J.~A., {Allende Prieto}, C., {et~al.} 2015, \aj,
  150, 173

\bibitem[{{Pancino} {et~al.}(2017){Pancino}, {Romano}, {Tang},
  {Tautvai{\v{s}}ien{\.{e}}}, {Casey}, {Gruyters}, {Geisler}, {San Roman},
  {Randich}, {Alfaro}, {Bragaglia}, {Flaccomio}, {Korn}, {Recio-Blanco},
  {Smiljanic}, {Carraro}, {Bayo}, {Costado}, {Damiani}, {Jofr{\'e}}, {Lardo},
  {de Laverny}, {Monaco}, {Morbidelli}, {Sbordone}, {Sousa}, \&
  {Villanova}}]{Pancino2017}
{Pancino}, E., {Romano}, D., {Tang}, B., {et~al.} 2017, \aap, 601, A112

\bibitem[{{Plez}(2012)}]{Plez2012}
{Plez}, B. 2012, {Turbospectrum: Code for spectral synthesis}

\bibitem[{{Press} {et~al.}(2002){Press}, {Teukolsky}, {Vetterling}, \&
  {Flannery}}]{Press2002}
{Press}, W.~H., {Teukolsky}, S.~A., {Vetterling}, W.~T., \& {Flannery}, B.~P.
  2002, {Numerical recipes in C++ : the art of scientific computing}

\bibitem[{{Ramos} {et~al.}(2020){Ramos}, {Mateu}, {Antoja}, {Helmi},
  {Castro-Ginard}, {Balbinot}, \& {Carrasco}}]{Ramos2020}
{Ramos}, P., {Mateu}, C., {Antoja}, T., {et~al.} 2020, \aap, 638, A104

\bibitem[{{Recio-Blanco} {et~al.}(2017){Recio-Blanco}, {Rojas-Arriagada}, {de
  Laverny}, {Mikolaitis}, {Hill}, {Zoccali}, {Fern{\'a}ndez-Trincado}, {Robin},
  {Babusiaux}, {Gilmore}, {Randich}, {Alfaro}, {Allende Prieto}, {Bragaglia},
  {Carraro}, {Jofr{\'e}}, {Lardo}, {Monaco}, {Morbidelli}, \&
  {Zaggia}}]{Recio-Blanco2017}
{Recio-Blanco}, A., {Rojas-Arriagada}, A., {de Laverny}, P., {et~al.} 2017,
  \aap, 602, L14

\bibitem[{{Renzini} {et~al.}(2015){Renzini}, {D'Antona}, {Cassisi}, {King},
  {Milone}, {Ventura}, {Anderson}, {Bedin}, {Bellini}, {Brown}, {Piotto}, {van
  der Marel}, {Barbuy}, {Dalessandro}, {Hidalgo}, {Marino}, {Ortolani},
  {Salaris}, \& {Sarajedini}}]{Renzini2015}
{Renzini}, A., {D'Antona}, F., {Cassisi}, S., {et~al.} 2015, \mnras, 454, 4197

\bibitem[{{Robin} {et~al.}(2003){Robin}, {Reyl{\'e}}, {Derri{\`e}re}, \&
  {Picaud}}]{Robin2003}
{Robin}, A.~C., {Reyl{\'e}}, C., {Derri{\`e}re}, S., \& {Picaud}, S. 2003,
  \aap, 409, 523

\bibitem[{{Schiavon} {et~al.}(2017){Schiavon}, {Zamora}, {Carrera},
  {Lucatello}, {Robin}, {Ness}, {Martell}, {Smith},
  {Garc{\'\i}a-Hern{\'a}ndez}, {Manchado}, {Sch{\"o}nrich}, {Bastian},
  {Chiappini}, {Shetrone}, {Mackereth}, {Williams}, {M{\'e}sz{\'a}ros},
  {Allende Prieto}, {Anders}, {Bizyaev}, {Beers}, {Chojnowski}, {Cunha},
  {Epstein}, {Frinchaboy}, {Garc{\'\i}a P{\'e}rez}, {Hearty}, {Holtzman},
  {Johnson}, {Kinemuchi}, {Majewski}, {Muna}, {Nidever}, {Nguyen}, {O'Connell},
  {Oravetz}, {Pan}, {Pinsonneault}, {Schneider}, {Schultheis}, {Simmons},
  {Skrutskie}, {Sobeck}, {Wilson}, \& {Zasowski}}]{Schiavon2017}
{Schiavon}, R.~P., {Zamora}, O., {Carrera}, R., {et~al.} 2017, \mnras, 465, 501

\bibitem[{{Shetrone} {et~al.}(2015){Shetrone}, {Bizyaev}, {Lawler}, {Allende
  Prieto}, {Johnson}, {Smith}, {Cunha}, {Holtzman}, {Garc{\'\i}a P{\'e}rez},
  {M{\'e}sz{\'a}ros}, {Sobeck}, {Zamora}, {Garc{\'\i}a-Hern{\'a}ndez}, {Souto},
  {Chojnowski}, {Koesterke}, {Majewski}, \& {Zasowski}}]{Shetrone2015}
{Shetrone}, M., {Bizyaev}, D., {Lawler}, J.~E., {et~al.} 2015, \apjs, 221, 24

\bibitem[{{Sills} {et~al.}(2019){Sills}, {Dalessandro}, {Cadelano},
  {Alfaro-Cuello}, \& {Kruijssen}}]{Sills2019}
{Sills}, A., {Dalessandro}, E., {Cadelano}, M., {Alfaro-Cuello}, M., \&
  {Kruijssen}, J.~M.~D. 2019, \mnras, 490, L67

\bibitem[{{Smith} {et~al.}(2013){Smith}, {Cunha}, {Shetrone}, {Meszaros},
  {Allende Prieto}, {Bizyaev}, {Garc{\'\i}a P{\'e}rez}, {Majewski}, {Schiavon},
  {Holtzman}, \& {Johnson}}]{Smith2013}
{Smith}, V.~V., {Cunha}, K., {Shetrone}, M.~D., {et~al.} 2013, \apj, 765, 16

\bibitem[{{Tang} {et~al.}(2018){Tang}, {Fern{\'a}ndez-Trincado}, {Geisler},
  {Zamora}, {M{\'e}sz{\'a}ros}, {Masseron}, {Cohen},
  {Garc{\'\i}a-Hern{\'a}ndez}, {Dell'Agli}, {Beers}, {Schiavon}, {Sohn},
  {Hasselquist}, {Robin}, {Shetrone}, {Majewski}, {Villanova}, {Schiappacasse
  Ulloa}, {Lane}, {Minnti}, {Roman-Lopes}, {Almeida}, \& {Moreno}}]{Tang2018}
{Tang}, B., {Fern{\'a}ndez-Trincado}, J.~G., {Geisler}, D., {et~al.} 2018,
  \apj, 855, 38

\bibitem[{{Ting} {et~al.}(2019){Ting}, {Conroy}, {Rix}, \&
  {Cargile}}]{Ting2019}
{Ting}, Y.-S., {Conroy}, C., {Rix}, H.-W., \& {Cargile}, P. 2019, \apj, 879, 69

\bibitem[{{Totten} \& {Irwin}(1998)}]{Totten1998}
{Totten}, E.~J. \& {Irwin}, M.~J. 1998, \mnras, 294, 1

\bibitem[{{Vasiliev} \& {Belokurov}(2020)}]{Vasiliev2020b}
{Vasiliev}, E. \& {Belokurov}, V. 2020, \mnras, 497, 4162

\bibitem[{{Vasiliev} {et~al.}(2020){Vasiliev}, {Belokurov}, \&
  {Erkal}}]{Vasiliev2020}
{Vasiliev}, E., {Belokurov}, V., \& {Erkal}, D. 2020, \mnras
  [\eprint[arXiv]{2009.10726}]

\bibitem[{{Villanova} {et~al.}(2016){Villanova}, {Monaco}, {Moni Bidin}, \&
  {Assmann}}]{Villanova2016}
{Villanova}, S., {Monaco}, L., {Moni Bidin}, C., \& {Assmann}, P. 2016, \mnras,
  460, 2351

\bibitem[{{Wilson} {et~al.}(2012){Wilson}, {Hearty}, {Skrutskie}, {Majewski},
  {Schiavon}, {Eisenstein}, {Gunn}, {Holtzman}, {Nidever}, {Gillespie},
  {Weinberg}, {Blank}, {Henderson}, {Smee}, {Barkhouser}, {Harding}, {Hope},
  {Fitzgerald}, {Stolberg}, {Arns}, {Nelson}, {Brunner}, {Burton}, {Walker},
  {Lam}, {Maseman}, {Barr}, {Leger}, {Carey}, {MacDonald}, {Ebelke}, {Beland},
  {Horne}, {Young}, {Rieke}, {Rieke}, {O'Brien}, {Crane}, {Carr}, {Harrison},
  {Stoll}, {Vernieri}, {Shetrone}, {Allende-Prieto}, {Johnson}, {Frinchaboy},
  {Zasowski}, {Garcia Perez}, {Bizyaev}, {Cunha}, {Smith}, {Meszaros}, {Zhao},
  {Hayden}, {Chojnowski}, {Andrews}, {Loomis}, {Owen}, {Klaene}, {Brinkmann},
  {Stauffer}, {Long}, {Jordan}, {Holder}, {Cope}, {Naugle}, {Pfaffenberger},
  {Schlegel}, {Blanton}, {Muna}, {Weaver}, {Snedden}, {Pan}, {Brewington},
  {Malanushenko}, {Malanushenko}, {Simmons}, {Oravetz}, {Mahadevan}, \&
  {Halverson}}]{Wilson2012}
{Wilson}, J.~C., {Hearty}, F., {Skrutskie}, M.~F., {et~al.} 2012, in Society of
  Photo-Optical Instrumentation Engineers (SPIE) Conference Series, Vol. 8446,
  Ground-based and Airborne Instrumentation for Astronomy IV, ed. I.~S.
  {McLean}, S.~K. {Ramsay}, \& H.~{Takami}, 84460H

\bibitem[{{Wilson} {et~al.}(2019){Wilson}, {Hearty}, {Skrutskie}, {Majewski},
  {Holtzman}, {Eisenstein}, {Gunn}, {Blank}, {Henderson}, {Smee}, {Nelson},
  {Nidever}, {Arns}, {Barkhouser}, {Barr}, {Beland}, {Bershady}, {Blanton},
  {Brunner}, {Burton}, {Carey}, {Carr}, {Colque}, {Crane}, {Damke}, {Davidson},
  {Dean}, {Di Mille}, {Don}, {Ebelke}, {Evans}, {Fitzgerald}, {Gillespie},
  {Hall}, {Harding}, {Harding}, {Hammond}, {Hancock}, {Harrison}, {Hope},
  {Horne}, {Karakla}, {Lam}, {Leger}, {MacDonald}, {Maseman}, {Matsunari},
  {Melton}, {Mitcheltree}, {O'Brien}, {O'Connell}, {Patten}, {Richardson},
  {Rieke}, {Rieke}, {Roman-Lopes}, {Schiavon}, {Sobeck}, {Stolberg}, {Stoll},
  {Tembe}, {Trujillo}, {Uomoto}, {Vernieri}, {Walker}, {Weinberg}, {Young},
  {Anthony-Brumfield}, {Bizyaev}, {Breslauer}, {De Lee}, {Downey}, {Halverson},
  {Huehnerhoff}, {Klaene}, {Leon}, {Long}, {Mahadevan}, {Malanushenko},
  {Nguyen}, {Owen}, {S{\'a}nchez-Gallego}, {Sayres}, {Shane}, {Shectman},
  {Shetrone}, {Skinner}, {Stauffer}, \& {Zhao}}]{Wilson2019}
{Wilson}, J.~C., {Hearty}, F.~R., {Skrutskie}, M.~F., {et~al.} 2019, \pasp,
  131, 055001

\bibitem[{{Yuan} {et~al.}(2020){Yuan}, {Chang}, {Beers}, \& {Huang}}]{Yuan2020}
{Yuan}, Z., {Chang}, J., {Beers}, T.~C., \& {Huang}, Y. 2020, \apjl, 898, L37

\bibitem[{{Zasowski} {et~al.}(2017){Zasowski}, {Cohen}, {Chojnowski},
  {Santana}, {Oelkers}, {Andrews}, {Beaton}, {Bender}, {Bird}, {Bovy},
  {Carlberg}, {Covey}, {Cunha}, {Dell'Agli}, {Fleming}, {Frinchaboy},
  {Garc{\'\i}a-Hern{\'a}ndez}, {Harding}, {Holtzman}, {Johnson}, {Kollmeier},
  {Majewski}, {M{\'e}sz{\'a}ros}, {Munn}, {Mu{\~n}oz}, {Ness}, {Nidever},
  {Poleski}, {Rom{\'a}n-Z{\'u}{\~n}iga}, {Shetrone}, {Simon}, {Smith},
  {Sobeck}, {Stringfellow}, {Szigeti{\'a}ros}, {Tayar}, \&
  {Troup}}]{Zasowski2017}
{Zasowski}, G., {Cohen}, R.~E., {Chojnowski}, S.~D., {et~al.} 2017, \aj, 154,
  198

\end{thebibliography}

\begin{appendix}
	
 \section{Spectrum of N-rich stars}
	
\begin{figure}	
		\begin{center}
			\includegraphics[width=90mm]{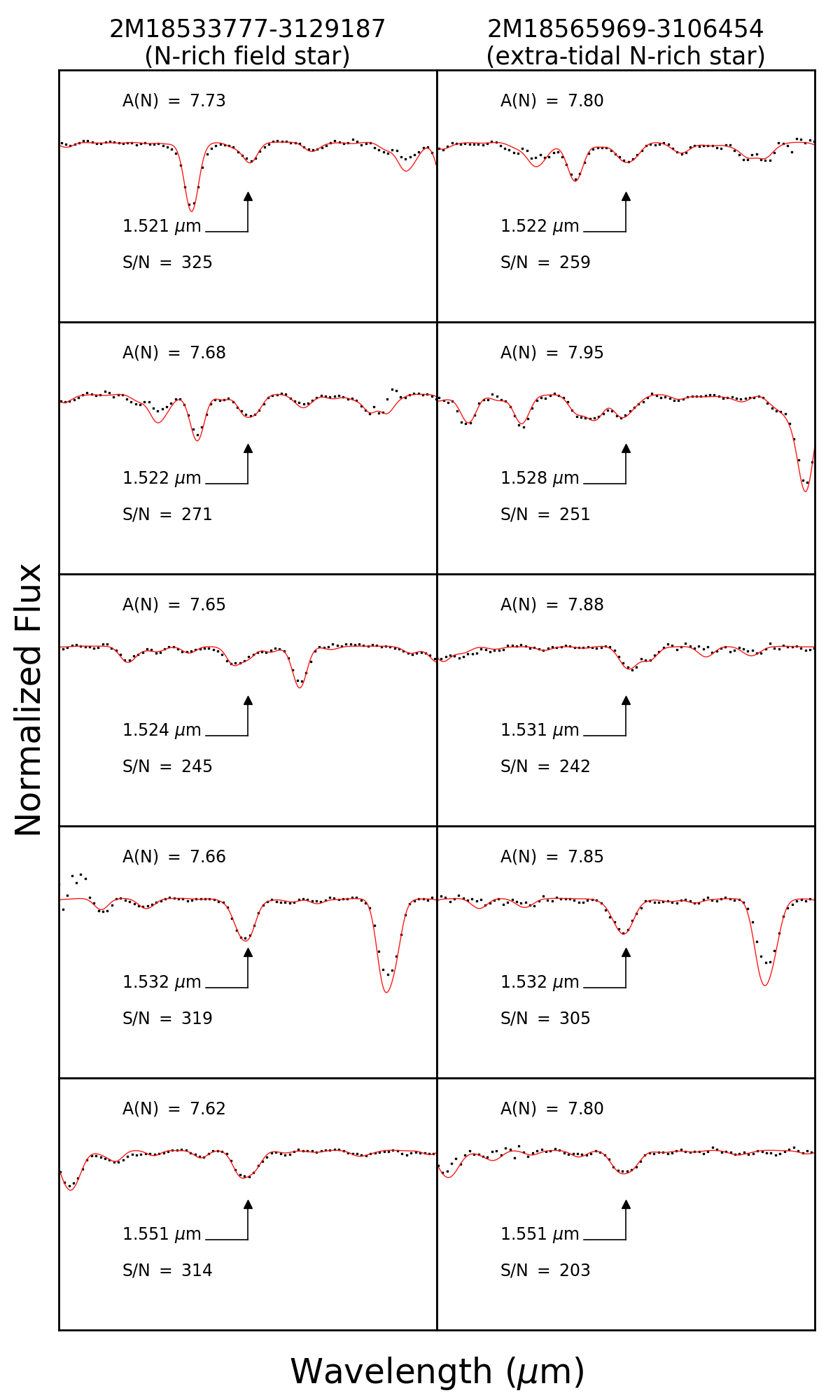}
			\caption{{\bf Detection of $^{12}$C$^{14}$N lines.} Spectral synthesis for the determination of nitrogen abundances for two N-rich stars beyond the tidal radius of M~54. Each panel shows the best-fit syntheses (red lines) from \texttt{BACCHUS} compared to the observed spectra (black squares) of selected $^{12}$C$^{14}$N lines (grey arrows)}
			\label{Figure4}
		\end{center}
	\end{figure}

Figure \ref{Figure4} shows an example of the local thermodynamic equilibrium (LTE) --\texttt{BACCHUS} spectral synthesis of selected $^{12}$C$^{14}$N lines for the two newly identified N-rich stars beyond the tidal radius of M~54. The black squares represent the observed spectrum, and the solid red line is the best abundance fit.

Table \ref{Table1} and \ref{Table11} list the atmospheric parameters, abundance ratios and the final errors in [Fe/H] and [X/Fe], while Table \ref{Table2} list of the main physical properties of our sample.

\begin{sidewaystable*}
			\begin{small}
	\setlength{\tabcolsep}{2.0mm}  
	\caption{Elemental abundances of our sample. }
	\centering
	\begin{tabular}{|l|ccccccccccccccccc|}
		\hline
		APOGEE ID   & $T_{eff}$ & $\log g $ & [M/H]   & $\xi_{t} $ & [C/Fe]      &   [N/Fe]    &   [O/Fe]     &    [Mg/Fe]     &    [Al/Fe]   &    [Si/Fe]    &  [K/Fe]      &   [Ca/Fe]     &  [Ti/Fe]    &  [Fe/H]     &  [Ni/Fe]      & [Ce/Fe]      &  [Nd/Fe]     \\    
		    & [K] & [cgs] &       & km s$^{-1}$ &      &      &       &         &    &       &        &      &    &    &      &       &     \\     		 
\hline 
{\bf  N-rich field star	}	    &&&       &  &      &      &       &         &    &       &        &      &    &    &      &       &     \\  
\hline
2M18533777$-$3129187 & 4158 & 1.23 & $-$1.35 & 1.8   & $-$0.60 & $+$1.21 &  $+$0.22 &  $+$0.21  & $+$0.58 &  $+$0.39 & $+$0.07 & $+$0.14  & $+$0.31 & $-$1.33 &  $-$0.01 & $+$0.22 & $+$0.30  \\ 
\hline 
{\bf   extra-tidal N-rich star	}	    &&&       &  &      &      &       &         &    &       &        &      &    &    &      &       &     \\     
\hline
2M18565969$-$3106454 & 4211 & 1.33 & $-$1.37 & 2.2   & $-$0.72 & $+$1.44 &  $+$0.19 &  $+$0.03  & $+$0.51 &  $+$0.26 & $+$0.13 & $+$0.11  & $+$0.28 & $-$1.39 &  $-$0.06 & $+$0.24 &     ...  \\     
\hline
   {\bf M~54 stars}	    &&&       &  &      &      &       &         &    &       &        &      &    &    &      &       &     \\     		 
\hline     
2M18544275$-$3029012 & 4480 & 1.08 & $-$1.40 & 1.8   &     ... &     ... &      ... &  $+$0.38  & $+$0.22 &  $+$0.23 &     ... & $+$0.32  & $+$0.17 & $-$1.37 &  $-$0.05 &     ... &     ...  \\          
2M18544848$-$3031588 & 4625 & 1.35 & $-$1.39 & 1.9   & $-$0.38 & $+$1.06 &  $+$0.96 &  $+$0.24  & $-$0.24 &  $+$0.26 & $+$0.10 & $+$0.23  & $+$0.48 & $-$1.42 &  $+$0.20 & $+$0.21 &     ...  \\          
2M18545303$-$3024220 & 4999 & 2.13 & $-$0.69 & 1.6   &     ... &     ... &      ... &  $+$0.29  & $-$0.08 &  $+$0.26 & $+$0.16 &     ...  &     ... & $-$1.32 &  $+$0.27 &     ... &     ...  \\          
2M18545434$-$3028458 & 5013 & 2.16 & $-$0.79 & 2.5   & $+$0.29 & $+$1.33 &  $+$1.30 &  $+$0.09  & $+$0.16 &  $+$0.26 & $+$0.16 &     ...  & $+$0.62 & $-$1.05 &  $-$0.01 &     ... &     ...  \\          
2M18545470$-$3027044 & 4271 & 0.69 & $-$1.31 & 2.6   &     ... & $+$0.59 &  $+$0.77 &  $+$0.17  & $+$0.13 &  $+$0.21 & $-$0.09 & $+$0.23  & $+$0.46 & $-$1.41 &  $-$0.02 & $+$0.28 &     ...  \\          
2M18545774$-$3025574 & 4110 & 0.37 & $-$1.37 & 2.9   &     ... & $+$0.35 &  $+$0.48 &  $+$0.03  & $+$0.02 &  $+$0.14 & $+$0.15 & $+$0.23  & $+$0.17 & $-$1.39 &  $+$0.05 & $+$0.03 &     ...  \\          
2M18545920$-$3028403 & 4360 & 0.85 & $-$1.42 & 2.0   & $-$0.43 & $+$1.00 &  $+$0.68 &  $+$0.16  & $+$0.24 &  $+$0.20 & $+$0.11 & $+$0.13  & $+$0.15 & $-$1.43 &  $+$0.04 & $+$0.17 & $+$0.44  \\          
2M18550076$-$3034068 & 4560 & 1.23 & $-$1.48 & 1.5   & $-$0.24 &     ... &  $+$0.76 &  $+$0.26  & $-$0.37 &  $+$0.24 &     ... & $+$0.23  & $+$0.10 & $-$1.39 &  $-$0.01 & $+$0.41 &     ...  \\          
2M18550247$-$3031314 & 4831 & 1.76 & $-$1.11 & 2.5   &     ... &     ... &      ... &  $+$0.13  & $-$0.17 &  $+$0.31 & $+$0.33 &     ...  &     ... & $-$1.32 &  $+$0.51 &     ... &     ...  \\          
2M18550293$-$3029523 & 4851 & 1.8  & $-$1.09 & 1.7   &     ... &     ... &      ... &  $+$0.22  &     ... &  $+$0.39 &     ... & $+$0.29  &     ... & $-$1.39 &  $+$0.28 &     ... &     ...  \\          
2M18550560$-$3026332 & 4477 & 1.07 & $-$0.54 & 2.5   & $-$0.14 & $+$2.20 &      ... &      ...  & $+$0.86 &  $+$0.43 & $+$0.44 &     ...  & $+$0.77 & $-$1.09 &  $-$0.04 &     ... &     ...  \\          
2M18550623$-$3030561 & 4584 & 1.28 & $-$1.11 & 2.3   & $-$0.24 & $+$1.48 &  $+$0.82 &  $+$0.22  & $+$0.54 &  $+$0.27 & $-$0.00 & $+$0.32  & $+$0.27 & $-$1.31 &  $+$0.00 & $+$0.38 &     ...  \\          
2M18550740$-$3026052 & 4233 & 0.62 & $-$1.32 & 2.0   & $-$0.51 & $+$1.00 &  $+$0.64 &  $+$0.24  & $-$0.21 &  $+$0.27 & $+$0.06 & $+$0.22  & $+$0.18 & $-$1.42 &  $+$0.09 &     ... &     ...  \\          
2M18550843$-$3029045 & 4238 & 0.63 & $-$0.88 & 2.7   & $-$0.27 & $+$1.89 &  $+$0.82 &  $+$0.29  & $+$0.41 &  $+$0.29 & $+$0.28 & $+$0.39  & $+$0.52 & $-$1.12 &  $+$0.02 &     ... &     ...  \\          
2M18551026$-$3033406 & 4110 & 0.37 & $-$1.36 & 2.9   & $-$0.72 & $+$1.35 &  $+$0.45 &  $+$0.04  & $+$0.36 &  $+$0.16 & $-$0.01 & $+$0.14  & $+$0.13 & $-$1.46 &  $+$0.01 & $+$0.12 &     ...  \\          
2M18551034$-$3027105 & 4463 & 1.05 & $-$1.14 & 1.7   & $-$0.48 & $+$0.99 &  $+$0.77 &  $+$0.20  & $-$0.12 &  $+$0.20 & $-$0.12 & $+$0.14  &    ...  & $-$1.21 &  $-$0.01 &     ... &     ...  \\          
2M18551061$-$3029597 & 4837 & 1.77 & $-$1.17 & 1.5   &     ... &     ... &      ... &  $+$0.11  & $+$0.63 &  $+$0.23 & $+$0.30 & $+$0.35  & $+$0.30 & $-$1.28 &  $+$0.19 &     ... &     ...  \\          
2M18551152$-$3025352 & 4961 & 2.05 & $-$1.38 & 1.7   &     ... &     ... &      ... &  $+$0.32  & $-$0.19 &  $+$0.25 &     ... & $+$0.33  & $+$0.50 & $-$1.22 &  $+$0.00 &     ... &     ...  \\          
2M18551416$-$3028548 & 4161 & 0.48 & $-$1.32 & 2.7   & $-$0.31 & $+$0.90 &  $-$0.40 &  $-$0.02  & $+$0.90 &  $+$0.24 & $+$0.21 & $+$0.26  & $+$0.26 & $-$1.27 &  $-$0.02 & $+$0.15 &     ...  \\          
2M18551535$-$3024142 & 4084 & 0.32 & $-$1.31 & 2.7   & $-$0.57 & $+$0.67 &  $+$0.53 &  $+$0.03  & $-$0.31 &  $+$0.15 & $+$0.11 & $+$0.20  & $+$0.08 & $-$1.29 &  $-$0.04 & $+$0.07 &     ...  \\          
2M18551672$-$3027508 & 4695 & 1.49 & $-$0.74 & 2.2   &    ...  & $+$1.28 &      ... &      ...  & $+$0.24 &  $+$0.55 & $+$0.54 &     ...  &     ... & $-$1.28 &  $+$0.17 &    ...  &     ...  \\          
2M18552982$-$3028545 & 4228 & 0.61 & $-$1.43 & 2.0   & $-$0.64 & $+$0.58 &  $+$0.41 &  $+$0.19  & $-$0.20 &  $+$0.12 & $-$0.01 & $+$0.22  & $+$0.13 & $-$1.18 &  $-$0.08 & $+$0.02 &     ...  \\    
		\hline
	\end{tabular}  \label{Table1}
		\end{small}
		\end{sidewaystable*}

\begin{sidewaystable*}
	\begin{small}
		\setlength{\tabcolsep}{2.0mm}  
		\caption{Final errors in [Fe/H] and [X/Fe].}
		\centering
		\begin{tabular}{|l|ccccccccccccc|}
			\hline
APOGEE ID            &  $\sigma_{\rm [C/Fe]}$ &  $\sigma_{\rm [N/Fe]}$ & $\sigma_{\rm [O/Fe]}$ &  $\sigma_{\rm [Mg/Fe]}$ &  $\sigma_{\rm [Al/Fe]}$ &  $\sigma_{\rm [Si/Fe]}$ &  $\sigma_{\rm [K/Fe]}$ &  $\sigma_{\rm [Ca/Fe]}$ &  $\sigma_{\rm [Ti/Fe]}$ & $\sigma_{\rm [Fe/H]}$ &  $\sigma_{\rm [Ni/Fe]}$ &  $\sigma_{\rm [Ce/Fe]}$ &  $\sigma_{\rm [Nd/Fe]}$  \\	
\hline		
{\bf  N-rich field star	}	     &&&       &  &      &      &       &         &    &       &        &      &         \\
\hline
2M18533777$-$3129187 &  0.03    &  0.06    & 0.17    &  0.07     &  0.04     &  0.06     &  0.04    &  0.11     &  0.12     & 0.05     &  0.11     &  0.08     &  0.04      \\
\hline 
{\bf   extra-tidal N-rich star	}	    &&&       &  &      &      &       &         &    &       &        &      &         \\     
\hline
2M18565969$-$3106454 &  0.03    &  0.08    & 0.20    &  0.07     &  0.06     &  0.06     &  0.04    &  0.15     &  0.11     & 0.05     &  0.11     &  0.09     &  ...       \\
\hline
{\bf M~54 stars}	       &&&       &  &      &      &       &         &    &       &        &      &         \\   		 
\hline   
2M18544275$-$3029012 &  ...     &   ...    &  ...    &  0.09     &  0.07     &  0.09     &  ...     &  0.16     &  0.12     & 0.07     &  0.12     &  ...      &  ...       \\
2M18544848$-$3031588 &  0.05    &  0.14    & 0.16    &  0.08     &  0.06     &  0.08     &  0.05    &  0.13     &  0.09     & 0.06     &  0.15     &  0.10     &  ...       \\
2M18545303$-$3024220 &  ...     &   ...    &  ...    &  0.10     &  0.08     &  0.21     &  0.08    &  ...      &  ...      & 0.09     &  0.11     &  ...      &  ...       \\
2M18545434$-$3028458 &  0.04    &  0.09    & 0.09    &  0.06     &  0.09     &  0.06     &  0.06    &  ...      &  0.12     & 0.04     &  0.11     &  ...      &  ...       \\
2M18545470$-$3027044 &  ...     &  0.10    & 0.17    &  0.08     &  0.08     &  0.07     &  0.08    &  0.07     &  0.07     & 0.05     &  0.07     &  0.11     &  ...       \\
2M18545774$-$3025574 &  ...     &  0.08    & 0.13    &  0.08     &  0.13     &  0.07     &  0.06    &  0.06     &  0.11     & 0.06     &  0.12     &  0.12     &  ...       \\
2M18545920$-$3028403 &  0.06    &  0.07    & 0.04    &  0.08     &  0.07     &  0.06     &  0.06    &  0.12     &  0.08     & 0.04     &  0.11     &  0.07     &  0.04      \\
2M18550076$-$3034068 &  0.06    &  ...     & 0.13    &  0.08     &  0.05     &  0.08     &  ...     &  0.12     &  0.09     & 0.06     &  0.12     &  0.06     &  ...       \\
2M18550247$-$3031314 &  ...     &  ...     & ...     &  0.09     &  0.07     &  0.13     &  0.07    &  ...      &  ...      & 0.08     &  0.11     &  ...      &  ...       \\
2M18550293$-$3029523 &  ...     &  ...     & ...     &  0.14     &  ...      &  0.22     &  ...     &  0.15     &  ...      & 0.09     &  0.08     &  ...      &  ...       \\
2M18550560$-$3026332 &  0.06    &  0.17    & ...     &  ...      &  0.07     &  0.07     &  0.07    &  ...      &  0.08     & 0.07     &  0.12     &  ...      &  ...       \\
2M18550623$-$3030561 &  0.06    &  0.08    & 0.06    &  0.08     &  0.07     &  0.05     &  0.06    &  0.21     &  0.09     & 0.07     &  0.08     &  0.07     &  ...       \\
2M18550740$-$3026052 &  0.10    &  0.08    & 0.15    &  0.09     &  0.10     &  0.12     &  0.05    &  0.13     &  0.13     & 0.06     &  0.12     &  ...      &  ...       \\
2M18550843$-$3029045 &  0.04    &  0.10    & 0.20    &  0.11     &  0.08     &  0.06     &  0.05    &  0.15     &  0.10     & 0.11     &  0.05     &  ...      &  ...      \\
2M18551026$-$3033406 &  0.07    &  0.07    & 0.07    &  0.07     &  0.14     &  0.05     &  0.07    &  0.11     &  0.07     & 0.12     &  0.12     &  0.08     &  ...       \\
2M18551034$-$3027105 &  0.07    &  0.08    & 0.05    &  0.07     &  0.09     &  0.06     &  0.05    &  0.09     &  ...      & 0.05     &  0.13     &  ...      &  ...       \\
2M18551061$-$3029597 &  ...     &  ...     & ...     &  0.08     &  0.11     &  0.08     &  0.07    &  0.07     &  0.13     & 0.06     &  0.13     &  ...      &  ...       \\
2M18551152$-$3025352 &  ...     &  ...     & ...     &  0.08     &  0.06     &  0.08     &  ...     &  0.12     &  0.10     & 0.07     &  0.12     &  ...      &  ...       \\
2M18551416$-$3028548 &  0.04    &  0.07    & 0.18    &  0.07     &  0.07     &  0.08     &  0.05    &  0.15     &  0.15     & 0.08     &  0.14     &  0.09     &  ...       \\
2M18551535$-$3024142 &  0.05    &  0.08    & 0.08    &  0.05     &  0.09     &  0.08     &  0.05    &  0.05     &  0.11     & 0.07     &  0.09     &  0.09     &  ...       \\
2M18551672$-$3027508 &  ...     &  0.11    & ...     &  ...      &  0.08     &  0.16     &  0.09    &  ...      &  ...      & 0.08     &  0.12     &  ...      &  ...       \\
2M18552982$-$3028545 &  0.06    &  0.05    & 0.09    &  0.15     &  0.07     &  0.08     &  0.06    &  0.08     &  0.07     & 0.05     &  0.09     &  0.08     &  ...       \\
			\hline
		\end{tabular}  \label{Table11}
	\end{small}
\end{sidewaystable*}

	\begin{table*}
			\setlength{\tabcolsep}{1.5mm}  
			\caption{Astrometric and kinematic properties of our sample. The last two columns indicate the typical S/N of the spectra and number of APOGEE visits.}
			\centering
			\begin{tabular}{|l|cccccccccc|}
				\hline
 APOGEE ID            & \texttt{RUWE} & $\mu_{\alpha}\cos(\delta) \pm \Delta$   & $\mu_{\delta} \pm \Delta$      & $G$     & $G_{BP}$   & $G_{RP}$   & $RV$     & $RV$-scatter & S/N & \# Visits \\         
	    && mas yr$^{-1}$ & mas yr$^{-1}$     & [mag]  &   [mag]    &  [mag]     &   km s$^{-1}$    &  km s$^{-1}$       &   pixel$^{-1}$ &        \\  
\hline 
{\bf  N-rich field star	}	    &&&       &  &      &      &       &         &   pixel$^{-1}$ &        \\  
\hline
2M18533777$-$3129187 & 0.9  &    0.001$\pm$0.016 & $-$4.150$\pm$0.013 & 12.79 & 13.62 & 11.91 & 181.02 & 0.26  & 402 & 6  \\ 
\hline 
{\bf   extra-tidal N-rich star	}	    &&&       &  &      &      &       &         &    &          \\     
\hline
2M18565969$-$3106454 & 0.9  & $-$2.701$\pm$0.027 & $-$1.336$\pm$0.023 & 14.75 & 15.59 & 13.86 & 134.75 & 0.55  & 225 & 12 \\   
\hline
{\bf M~54 stars}	    &&&       &  &      &      &       &         &    &      \\     		 
\hline            
2M18544275$-$3029012 & 1.0  & $-$2.667$\pm$0.040 & $-$1.357$\pm$0.033 & 15.55 & 16.27 & 14.71 & 137.07 & 0.33  & 51  & 4  \\          
2M18544848$-$3031588 & 1.1  & $-$2.686$\pm$0.037 & $-$1.356$\pm$0.030 & 15.17 & 15.91 & 14.32 & 141.74 & 0.34  & 172 & 12 \\          
2M18545303$-$3024220 & 0.8  & $-$2.584$\pm$0.052 & $-$1.430$\pm$0.045 & 16.27 & 16.91 & 15.49 & 147.14 & 0.91  & 53  & 10 \\          
2M18545434$-$3028458 & 1.1  & $-$2.694$\pm$0.052 & $-$1.440$\pm$0.045 & 15.72 & 16.33 & 14.85 & 142.11 & 0.42  & 109 & 12 \\          
2M18545470$-$3027044 & 1.4  & $-$2.412$\pm$0.035 & $-$1.345$\pm$0.030 & 14.60 & 15.44 & 13.68 & 131.88 & 1.16  & 266 & 12 \\          
2M18545774$-$3025574 & 1.1  & $-$2.717$\pm$0.029 & $-$1.321$\pm$0.025 & 14.68 & 15.54 & 13.75 & 148.86 & 1.76  & 262 & 12 \\          
2M18545920$-$3028403 & 1.1  & $-$2.800$\pm$0.040 & $-$1.374$\pm$0.035 & 14.88 & 15.54 & 13.92 & 145.79 & 0.71  & 217 & 12 \\          
2M18550076$-$3034068 & 0.9  & $-$2.685$\pm$0.036 & $-$1.494$\pm$0.030 & 15.35 & 16.07 & 14.52 & 149.07 & 0.38  & 118 & 6  \\          
2M18550247$-$3031314 & 1.0  & $-$2.625$\pm$0.049 & $-$1.310$\pm$0.040 & 16.10 & 16.76 & 15.33 & 138.48 & 0.77  & 82  & 12 \\          
2M18550293$-$3029523 & 1.0  & $-$2.526$\pm$0.045 & $-$1.230$\pm$0.037 & 15.82 & 16.30 & 14.89 & 141.96 & 0.49  & 115 & 12 \\          
2M18550560$-$3026332 & 0.9  & $-$2.617$\pm$0.045 & $-$1.345$\pm$0.038 & 15.93 & 16.60 & 15.12 & 143.86 & 0.29  & 93  & 10 \\          
2M18550623$-$3030561 & 1.0  & $-$2.832$\pm$0.038 & $-$1.366$\pm$0.032 & 15.57 & 16.29 & 14.74 & 135.17 & 0.31  & 136 & 12 \\          
2M18550740$-$3026052 & 1.0  & $-$2.658$\pm$0.028 & $-$1.243$\pm$0.023 & 14.82 & 15.65 & 13.94 & 132.96 & 0.95  & 93  & 4  \\          
2M18550843$-$3029045 & 1.0  & $-$2.557$\pm$0.038 & $-$1.340$\pm$0.031 & 15.46 & 16.16 & 14.54 & 135.85 & 0.35  & 158 & 12 \\          
2M18551026$-$3033406 & 1.1  & $-$2.644$\pm$0.027 & $-$1.407$\pm$0.021 & 14.49 & 15.41 & 13.54 & 142.01 & 0.74  & 309 & 12 \\          
2M18551034$-$3027105 & 1.0  & $-$2.682$\pm$0.037 & $-$1.379$\pm$0.030 & 15.44 & 16.19 & 14.60 & 154.11 & 0.31  & 106 & 6  \\          
2M18551061$-$3029597 & 1.1  & $-$2.888$\pm$0.054 & $-$1.325$\pm$0.044 & 16.07 & 16.65 & 15.24 & 147.11 & 0.94  & 78  & 11 \\          
2M18551152$-$3025352 & 1.0  & $-$2.665$\pm$0.053 & $-$1.407$\pm$0.044 & 16.22 & 16.74 & 15.35 & 146.85 & 0.62  & 73  & 10 \\          
2M18551416$-$3028548 & 1.1  & $-$2.675$\pm$0.031 & $-$1.342$\pm$0.026 & 14.98 & 15.80 & 14.10 & 139.67 & 0.38  & 208 & 12 \\          
2M18551535$-$3024142 & 1.0  & $-$2.683$\pm$0.029 & $-$1.403$\pm$0.023 & 14.74 & 15.62 & 13.82 & 151.11 & 1.01  & 230 & 12 \\          
2M18551672$-$3027508 & 0.9  & $-$2.692$\pm$0.043 & $-$1.407$\pm$0.035 & 15.80 & 16.44 & 15.00 & 140.11 & 0.94  & 76  & 11 \\          
2M18552982$-$3028545 & 0.9  & $-$2.689$\pm$0.033 & $-$1.292$\pm$0.028 & 15.15 & 15.96 & 14.27 & 135.09 & 0.38  & 192 & 12 \\    
				\hline
			\end{tabular}  \label{Table2}
	\end{table*}

\end{appendix} 
\end{document}